\documentstyle[11pt,epsf,rotate]{article}
\newcommand{\bce}{\begin{center}}
\newcommand{\ece}{\end{center}}
\newcommand{\be}{\begin{equation}}
\newcommand{\ee}{\end{equation}}
\newcommand{\bea}{\begin{eqnarray}}
\newcommand{\eea}{\end{eqnarray}}
\newcommand{\bdes}{\begin{description}}
\newcommand{\edes}{\end{description}}
\newcommand{\bit}{\begin{itemize}}
\newcommand{\eit}{\end{itemize}}
\newcommand{\E}{\>=\>}
\newcommand{\EA}{&=&}

\newcommand{\deF}{\> = \, : \>}
\newcommand{\Def}{\> : \, = \>}
\newcommand{\To}{\> \longrightarrow \> }
\newcommand{\bfl}{\begin{flushright}}
\newcommand{\efl}{\end{flushright}}

\newcommand{\non}{\nonumber \\}
\newcommand{\fp}{{\bf p}}
\newcommand{\fk}{{\bf k}}
\newcommand{\fx}{{\bf x}}
\newcommand{\fb}{{\bf b}}

\renewcommand{\thesection}{\Roman{section}.}

\begin{document}

\thispagestyle{empty}
\bfl
PSI-PR-08-06
\efl

\vspace*{2cm}

\bce

{\Large Perturbation Theory Without Diagrams: The Polaron Case}

\vspace*{2cm}
R. Rosenfelder 

\vspace{0.6cm}
Particle Theory Group, Paul Scherrer Institute, CH-5232 Villigen PSI, 
Switzerland 
\ece

\vspace*{4cm}

\begin{abstract}
\noindent
Higher-order perturbative calculations in Quantum
(Field) Theory suffer from the factorial increase of the number of
individual diagrams. Here I describe an approach which evaluates
the total contribution numerically for finite temperature
from the cumulant expansion of the corresponding observable followed
by an extrapolation to zero temperature.
This method (originally proposed by Bogolyubov and Plechko) is applied
to the calculation of higher-order terms for the
ground-state energy of the polaron. 
Using state-of-the-art multidimensional integration routines two new coefficients
are obtained corresponding to a four- and five-loop
calculation. Several analytical and numerical procedures have been implemented which 
were crucial for obtaining reliable results.
\end{abstract}

\vspace{2cm}


\newpage
\setcounter{equation}{0}

\section{Introduction}
\label{sec: intro}
Highly accurate measurements require precise theoretical calculations 
which perturbation theory can yield if the coupling constant is small. However, in
 Quantum Field Theory  the number of diagrams grows factorially with the order of 
perturbation theory and they become more
and more complicated as the corresponding loop diagrams involve high-dimensional integrals 
over complicated (and singular) functions.

The prime example is the anomalous magnetic moment of the electron where
new experiments \cite{Han,Odo} need high-order quantum-electrodynamical 
calculations. In fact, the estimate for the fifth-order contribution is the largest source
of theoretical uncertainty if one attributes an  ``error'' to it at all \cite{Gab}. In 
addition, further improvements of the experimental accuracy are foreseen.

As derived in the textbook \cite{ItZu} the number of diagrams contributing to the vertex 
function in Quantum Electrodynamics (QED) is given by the coefficients of the generating function
\be
\Gamma \E \frac{4 z (1 - S)}{S^3} \> , \hspace{0.5cm} 
S \E - 2 z \left [ 1 + \frac{K_0'(z)}{K_0(z)} \right ]
\ee
[with $ z = - 1/(4 \alpha) $ and $K_0(z) $ the zeroth-order modified Bessel function 
of second kind] when expanded in powers of the fine-structure constant $\alpha$
\be
\Gamma(\alpha) \E 1 + \alpha + 7 \, \alpha^2 + 72 \, \alpha^3 + 891 \, \alpha^4 + 
12672 \, \alpha^5 + 202770 \, \alpha^6 + \ldots \> \> .
\ee
The contributions up to third order are known analytically \cite{LaRe} and the 891 diagrams
in fourth order have been evaluated numerically by Kinoshita and coworkers \cite{Kino4}. 
In view of the ever more precise experiments there 
are ongoing efforts \cite{Kino5} to calculate 
all $12 672$ diagrams in ${\cal O}(\alpha^5)$ numerically and by automated routines. This is 
a huge, heroic effort considering
the complexity of individual diagrams, the large cancellations among them and  
the intricacies of infrared and ultraviolet 
divergencies in the integrands.

Obviously new and more efficient methods would be most welcome for a cross-check
as well as further progress. However, it is useful first to consider a simpler field theory 
which is nontrivial but free from ultraviolet divergencies. This is supplied by the 
polaron problem -- the field theory of a single nonrelativistic electron slowly 
moving in a polarizable crystal and thereby
interacting with an infinite number of phonons. Similar as in Quantum Electrodynamics there
exists a large number of perturbative calculations
for the ground-state energy and other properties of the quasiparticle
which is made up by the electron and its surrounding cloud of virtual phonons.

In this paper we investigate a method originally proposed by 
Bogolyubov (Jr.) and Plechko (BP) \cite{BoPl} to obtain 
higher-order terms in the ground-state energy of a polaron without evaluating diagrams. 
As the polaron problem is the prototype of the worldline approach to relativistic Quantum Field Theory
\cite{WC1,WC2,QED1,QED2} we believe that a similar method also holds promise for high-order 
perturbative calculations in particle physics, in particular QED. 

Preliminary results have already been presented in Ref. \cite{Dresden}. Here I give a detailed account 
of the analytical and numerical methods which are required so that the BP method works. The paper 
is organized as follows.In Secs. \ref{sec: polaron} and \ref{sec: BP} we recall the basics of the
polaron model and the BP method. Section \ref{sec: tricks} gives an account of the necessary steps 
to obtain reliable numerical results. These are presented and discussed in Sec. \ref{sec: num}.
The last section contains our conclusion and the outlook for further work whereas more technical details
are collected in three appendices.

\section{The polaron problem - a nonrelativistic field theory}
\label{sec: polaron}

A model
Hamiltonian describing the dressing of the bare electron by a cloud of phonons
has been given by H. Fr\"ohlich
\be
\hat H \E \frac{1}{2} \hat \fp^2 + \int d^3 k \, \hat a^{\dagger}_{\fk} \, \hat a_{\fk}
+ i \left ( 2 \sqrt{2} \pi \alpha \right )^{1/2} \int \frac{d^3k}{(2 \pi)^3} \> \frac{1}{|\fk|} \, 
\left [ \, 
\hat a^{\dagger}_{\fk} \, e^{i \fk \cdot \hat \fx} - H.c. \, \right ] \> \> , \> \> 
\left [  \hat a_{\fk}, \hat a^{\dagger}_{\fk'} \right ] \E \delta^{(3)}(\fk - \fk')
\ee
where $\alpha$ is the dimensionless electron-phonon coupling constant.
Due to its interaction with the medium
the energy of the quasiparticle is changed and it acquires an effective mass
\be
E_{\fp} = E_0 + \frac{\fp^2}{2 m^{\star}}  + \ldots \> \> .
\ee
The aim is to calculate the power series expansion for the ground-state energy of a non-moving
polaron 
\be
E_0(\alpha) \deF \sum_{n=1} e_n \, \alpha^n
\label{def en}
\ee
as function of $\alpha$ \cite{fn_1}.
The lowest-order coefficients are well known
\bea
e_1 \EA -1 \\
\label{e1}
e_2 \EA  \frac{1}{\sqrt{2}} - \ln \left ( 1 + \frac{3}{4} \sqrt{2} \right ) 
\E -0.015 919622     \hspace{1cm}        \mbox{from Ref. \cite{HoMu}} \> ,
\label{e2} \\
e_3 \EA -0.000 806070     \hspace{5cm}        \mbox{from refs. \cite{Smon,SeSm} }  \> ,
\label{e3}
\eea
but there has been no progress towards higher-order terms. 

\vspace{0.2cm}

In the path-integral approach \cite{Feyn} the (infinite) phonon degrees of freedom may be
integrated out exactly which leads to an effective, two-time action
\be
S[\fx] \E \int_0^{\beta} dt \> \frac{1}{2}\, \dot \fx^2   - \frac{\alpha}{2 \sqrt{2}} \,
\int_0^{\beta}dt \, dt'  \> \, \frac{\cosh \left [\beta/2-|t-t'|) \right ]}{\sinh(\beta/2)} \>
\frac{1}{|\fx(t) - \fx(t')|} \> .
\label{S_pol1}
\ee
Here $\beta$ is the Euclidean time or inverse temperature. Some simplifications are possible: first,
the symmetry between the two times $t, t'$ allows us to restrict the integration range of the latter
to $ 0 \le t' \le t $ together with doubling the strength of the interaction. Second, as we are only 
interested in the ground-state energy $E_0$ of the polaron which can be obtained by the 
large-$\beta$ limit of the partition function
\be
Z \Def \int d^3x \int_{\fx(0) = \fx(\beta) = \fx} {\cal D}^3 x \> e^{-S[\fx]} \> \> 
\stackrel{\beta \to \infty}{\longrightarrow} \> \> \rm {const} \> e^{- \beta \, E_0}
\label{def partfunc} \> ,
\ee
we may replace
\be
 \frac{\cosh \left [ \beta/2-(t-t') \right ]}{\sinh(\beta/2)} \E \frac{\exp(-\sigma) +
\exp[-(\beta - \sigma)]}{1 - \exp(-\beta)}  \> \stackrel{\beta \to \infty}{\longrightarrow} \> 
\exp(-\sigma) \> ,
\ee
where $ \sigma = t - t' $ is the relative time \cite{fn_2}.
Thus, in the following, we will use
\be
S[\fx] \E \int_0^{\beta} dt \> \frac{1}{2}\, \dot \fx^2   - \frac{\alpha}{\sqrt{2}} \,
\int_0^{\beta}dt   \int_0^t dt' 
\frac{\exp(-\sigma)}{|\fx(t) - \fx(t')|} \deF S_0 + S_1 
\label{S_pol2}
\ee
as a full polaron action.

\vspace{0.3cm}

Useful order-of-magnitude estimates for higher-order energy coefficients 
can be obtained in various approximate
treatements of the polaron problem. Most prominent and successful among these
is Feynman's approach \cite{Feyn} in which a quadratic trial action 
\be
S_t \E \int_0^{\beta} dt \> \frac{1}{2} \dot \fx^2 + \int_0^{\beta} dt \int_0^t dt' \> f(t-t') \, 
\left [ \, \fx(t) - \fx(t') \, \right ]^2 
\ee
is used as variational approximation for the full action (\ref{S_pol2}). Feynman chose
an exponential form of the retardation function with
two variational parameters which are determined by minimizing Jensen's inequality.
The corresponding energy coefficients can be
calculated analytically to high order \cite{SeSm} as sketched in Appendix \ref{app: best}. 
The result is
\bea
e_1^{F} \EA -1 \> , \hspace{0.3cm} e_2^{F} \E -\frac{1}{81} \E - 1.234568 \times 10^{-2} \> ,
\hspace{0.3cm} e_3^{F} \E \frac{16}{729} - \frac{56}{6561} \sqrt{7} \E - 0.634366  \times 10^{-3} 
\> , \non
e_4^{F} \EA \frac{3200 \sqrt{10} -633236}{1594323} + \frac{78496}{531441} \sqrt{7} \E
-0.464315  \times 10^{-4} \> ,
\label{e4 Fey}\\
e_5^{F} \EA \frac{1673496632 -6044800 \sqrt{10}-70304 \sqrt{13}}{129140163} +
\frac{793600}{43046721} \sqrt{70}
- \frac{1476371144}{301327047} \sqrt{7} \non
\EA -0.395686 \times 10^{-5} \> .
\label{e5 Fey}
\eea

\noindent
However, one can do better by allowing the variational principle to 
determine the best retardation function itself. Then one gets \cite{best,shadow}
\be
e_1^{\rm best} \E - 1  \> , \hspace{0.3cm} e_2^{\rm best} \E -\left ( \frac{1}{12} - \frac{2}{9 \pi} 
\right ) \E -1.2597803 \times 10^{-2} \> . 
\ee
Note that $e_2^{\rm best}$ is only slightly better than $e_2^{F}$ despite the fact that the  
retardation function in the unrestricted variational approach has quite a different small-time
behavior than Feynman's parametrization. This is due to the (relative) insensitivity of the 
polaron energy to small-time dynamics. In this respect four-dimensional field theories in the worldline 
description are quite different, in particular realistic, renormalizable ones similar to  
QED \cite{QED2}.
Appendix \ref{app: best} also describes how one can obtain numerically the higher-order energy 
coefficients for the best
quadratic approximation. We have obtained the values
\be
e_3^{\rm best} \E -0.64650 \times 10^{-3} \>, \hspace{0.3cm} e_4^{\rm best} \E -0.4686 \times 10^{-4} 
\> ,
\hspace{0.3cm} e_5^{\rm best} \E -0.3940 \times 10^{-5} 
\ee
which -- again -- are not very much different from the results using the much simpler Feynman 
parametrization.

\section{The Bogoliubov-Plechko (BP) method}
\label{sec: BP}

In order to get the perturbative expansion of $E_0(\alpha) $ we use
the {\it cumulant} expansion of the partition
function for large $\beta$
\be
Z \E Z_0 \, \exp \left [  \, \sum_{n=1} \frac{(-)^n}{n !} \lambda_n(\beta) \, \right]
\ee
where $ \lambda_n(\beta) $ are the cumulants 
with respect to $ S_1 $ and $Z_0$ is
the free partition function for a system confined in a large volume. 

\vspace{0.2cm}
The cumulants 
(or semi-invariants)
are obtained from the (normalized) {\it moments}
\be
m_n \> \equiv \>  \left < S_1^n \right > \Def  C \, \int d^3x
\int_{\fx(0) = \fx}^{\fx(\beta) = \fx} {\cal D}^3 x \> \> 
S_1^n \>  e^{-S_0[\fx]} 
\ee
(here the $\beta$ dependence is suppressed and the normalization constant $C$ is chosen 
such that $m_0 = 1$) via the recursion relation
\be
\lambda_{n+1} \E m_{n+1}  -  \sum_{k=0}^{n-1}
{n \choose k} \lambda_{k+1}  \> m_{n-k}  \> .
\label{cum recurs}
\ee
This is standard and easily proved by differentiating the characteristic function 
\cite{quasi}
\be
\Phi(t) \E \left < e^{-t S_1} \right > \E \sum_{n=0} (-)^n \frac{t^n}{n !} \, m_n
\E \exp \left [ \, \sum_{n=1} (-)^n \frac{t^n}{n !} \, \lambda_n \, \right ]
\ee
with respect to $t$ in moment and cumulant form
\be
- \sum_{n=0} \frac{(-t)^n}{n !} \, m_{n+1} \E  -
\Phi(t) \, \cdot \, \sum_{n=0} \frac{(-t)^n}{n !} \, \lambda_{n+1} \> .
\ee
If the moment expansion for $\Phi(t)$ is inserted on the right-hand side one obtains after
rearrangement
\be
- \sum_{n=0} \frac{(-t)^n}{n !} \, m_{n+1}\E
-  \sum_{n=0} (-t)^n \, \sum_{k=0}^n \frac{1}{k ! (n-k) !} \lambda_{k+1} \, m_{n-k}
\ee
for all powers of $t$ which establishes Eq. (\ref{cum recurs}).
The first few cumulants are 
\bea
\lambda_1 \EA m_1  \\
\lambda_2 \EA m_2 -  m_1^2
\label{cum2}\\
\lambda_3 \EA m_3 - 3 \, m_2 \, m_1 + 2 \, m_1^3
\label{cum3}\\
\lambda_4 \EA m_4 - 4 \, m_3 \, m_1 - 3 \, m_2^2 + 12 \, m_2\,  m_1^2 - 6 \, m_1^4
\label{cum4}\\
\lambda_5 \EA m_5 - 5 \, m_4 \, m_1 - 10 \, m_3 \, m_2 + 20 \, m_3 \, m_1^2 +
30 \, m_2^2 \, m_1 - 60 \, m_2 \, m_1^3 + 24 \, m_1^5 \> \> .
\label{cum5}
\eea
For large $\beta$ we then get the ground-state energy as zero-temperature limit
of the free energy
\be
E_0 \E \lim_{\beta \to \infty} \left ( - \frac{1}{\beta} \right )
\sum_{n=1} \frac{(-)^n}{n !} \lambda_n(\beta)
\label{e0 by cum}
\ee
since the free partition function does not contribute. By construction the
$n$th moment is proportional to $\alpha^n$ and Eq. (\ref{cum recurs}) (and the examples)
show that the cumulants share this properties.
Comparing with Eq. (\ref{def en}) we see that
\be
e_n \E \frac{(-)^{n+1}}{\alpha^n \, n !} \, \lim_{\beta \to \infty} \frac{1}{\beta} \,
\lambda_n(\beta)  \> .
\label{en limit}
\ee

{\bf The moments $m_n$}. 
We calculate the moments $m_n$ by expanding the paths in Fourier components
\be
\fx(t) \E \sqrt{2 \beta} \, \fb_0 \, \frac{t}{\beta} + \sum_{k=1}^{\infty}
\frac{2 \sqrt{\beta}}{k \pi} \, \fb_k \, \sin \left (\frac{k \pi t}{\beta} \right )\> , \hspace{0.3cm}
\fx \deF \sqrt{2 \beta} \, \fb_0
\ee
so that
\be
S_0 \E \sum_{k=0}^{\infty} \fb_k^2
\ee 
and the functional integration is over the coefficients $\fb_k , \> k = 0, 1, \ldots $. 
Writing
\be
S_1 \E - \frac{\alpha}{\sqrt{2}} \, \int_0^{\beta} dt \int_0^t dt' \, e^{-(t-t')}
\int \frac{d^3 p}{2 \pi^2} \> \frac{1}{\fp^2} \, \exp \left \{ i  \fp \cdot \left [
\fx(t) - \fx(t') \right ] \, \right \}
\ee
we have
\bea
m_n \EA (-)^n \frac{\alpha^n}{2^{n/2}} \,  \int_0^{\beta} dt_1 \ldots dt_n \,
\int_0^{t_1} dt_1' \ldots \int_0^{t_n} dt_n' \> \exp \left [ - (t_1 - t_1') - \ldots
-(t_n - t_n') \right ] \non
&& \times \int \frac{d^3 p_1}{2 \pi^2} \> \frac{1}{\fp_1^2} \ldots
\int \frac{d^3 p_n}{2 \pi^2} \> \frac{1}{\fp_n^2} \> \left < \, \exp \left [ 2i
\sum_{m=1}^n \fp_m \cdot \sum_{k=0}^{\infty} \ell_k(t_m,t_m') \, \fb_k \right ] \,
\right >
\eea
where
\be
 \ell_k(t,t') \E \left \{ \begin{array}{l@{\quad:\quad}l}
                          \frac{1}{\sqrt{2 \beta}} (t-t') & k = 0 \, , \\
                           \frac{\sqrt{\beta}}{k \pi} \left ( \sin \frac{k \pi t}{\beta}
                              - \sin \frac{k \pi t'}{\beta} \right ) & k \ge 1 
                           \end{array} \right.
\ee
and 
\be
\left < \, {\cal O} \, \right > \Def \frac{\int d^3b_0 \, d^3b_1 \ldots \> 
{\cal O}(\fb_0,\fb_1 \ldots) \, 
\exp\left [-S_0 (\fb_0,\fb_1 \ldots ) \right ]}{\int d^3b_0 \, d^3b_1 \ldots \> 
\exp \left [-S_0 (\fb_0,\fb_1 \ldots ) \right ] }
\ee
is the average with respect to the free action $S_0$.
\vspace{0.1cm}

As a Gaussian integral over the $\fb_k$'s this average  can be done easily and
one obtains
\bea
m_n \EA (-)^n \frac{\alpha^n}{2^{n/2}} \,  \prod_{m=1}^n \left ( \int_0^{\beta} dt_m
\int_0^{t_m} dt_m' \right )\> \exp \left [ - \sum_{m=1}^n (t_m - t_m') \right ] \non
&& \times  \prod_{m=1}^n  \left ( \int \frac{d^3 p_m}{2 \pi^2} \> \frac{1}{\fp_m^2}
\right ) \> \exp \left [ - \sum_{k=0}^{\infty} \left (
\sum_{m=1}^n \ell_k(t_m,t_m') \, \fp_m \right )^2 \right ] \> .
\eea
If we now write the  $m$th Coulomb propagator as
\be
\frac{1}{\fp_m^2} \E \frac{1}{2} \,
\int_0^{\infty} du_m \> \exp \left [ - \frac{1}{2} \fp_m^2 \, u_m \right ]
\ee
then {\it all} momentum integrations can be performed and give the result
\bea
m_n \EA (-)^n \frac{\alpha^n}{(4 \pi)^{n/2}} \, \prod_{m=1}^n \left ( \int_0^{\beta}
dt_m \int_0^{t_m} dt_m' \int_0^{\infty} du_m \right ) \, \exp \left [ - \sum_{m=1}^n
(t_m - t_m') \right ] \non
&& \hspace{3cm} \times  \left [ {\rm det}_n \, A \left (t_1, \ldots , t_n, t_1', \ldots ,  t_n';
u_1, \ldots , u_n \right ) \right ]^{-3/2} \> .
\label{m_n}
\eea
Here $(A)$ is the $n \times n $ matrix made up by the elements
\be
(A)_{i j} \E 2 \sum_{k=0}^{\infty} \ell_k(t_i,t_i') \ell_k(t_j,t_j') + u_i \, \delta_{ij}
\deF a_{ij} + u_i \, \delta_{ij} \> .
\label{(A)}
\ee
It is essential that the infinite sum over the modes $k$ can be performed
analytically. Using Eq. 1.443.3 in Ref. \cite{GR}
\be
\sum_{k=1}^{\infty} \, \frac{\cos kx}{k^2} \E \frac{\pi^2}{6} - \frac{\pi x}{2}
+ \frac{x^2}{4}\> ,  \hspace{1cm} 0 \le x \le 2 \pi
\ee
we indeed have
\bea
\sum_{k=1}^{\infty} \frac{1}{k^2 \pi^2} \, \sin \frac{k \pi x}{\beta} \,
\sin \frac{k \pi y }{\beta} \EA \frac{1}{2} \sum_{k=1}^{\infty} \frac{1}{k^2 \pi^2} \,
\left [ \, \cos \frac{k \pi (x-y)}{\beta} - \cos \frac{k \pi (x+y)}{\beta} \, \right ]
\non
\EA \frac{1}{2 \beta} \left [ \, {\rm min}(x,y) - \frac{x y}{\beta} \, \right ] \> ,
\quad 0 \le x,y \le \beta
\eea
and, therefore,
\be
a_{ij} \E {\rm min}(t_i,t_j) -
{\rm min}(t_i,t_j') - {\rm min}(t_i',t_j) + {\rm min}(t_i',t_j') \> .
\ee
Using $ \> {\rm min}(x,y) = \left [ \> x + y - |x - y| \> \right ]/2 \> $
this may also be  written as
\be
a_{ij} \E  \frac{1}{2}  \left [ \, - |t_i -t_j| + |t_i - t_j'| + | t_i' - t_j| -
|t_i' - t_j'| \, \right ] \> .
\label{a_ij}
\ee
Note that $a_{ij} = a_{ji}$ and that
\be
a_{ii} \E t_i - t_i' \deF \sigma_i \> \ge 0 
\ee
since $ t_i \ge t_i'$ .
This is a special case of the more general fact that $(A)$ is a positive
definite matrix (otherwise the momentum integral would not converge) \cite{fn_3}.
Well-known theorems of matrix analysis \cite{HoJo,HoSt} then guarantee
that the principal minors of all orders are non-negative and the diagonal elements
are just the ones of lowest order. 
\vspace{0.2cm}

Introducing total and relative times
\be
\sigma_i \Def t_i - t_i'  \> \> , \hspace{0.3cm} \Sigma_i \Def \frac{t_i + t_i'}{2}
\label{def sigma}
\ee
we have
\bea
m_n \EA (-)^n \frac{\alpha^n}{(4 \pi)^{n/2}} \, \prod_{m=1}^n \left ( \int_0^{\beta}
d\sigma_m \int_{\sigma_m/2}^{\beta - \sigma_m/2} d\Sigma_m \int_0^{\infty} du_m \right ) \, 
\exp \left [ - \sum_{m=1}^n \sigma_m \right ] \non
&& \hspace{3cm} \times  \left [ {\rm det}_n \, A \left (\sigma_1, \ldots , \sigma_n, \Sigma_1, 
\ldots , \Sigma_n; u_1,  \ldots , u_n \right ) \right ]^{-3/2} \> .
\label{m_n 2}
\eea
Due to time-translational invariance, 
the nondiagonal matrix elements, say $a_{12}$, only depend on three variables which we denote by
\be
S \Def \Sigma_1 - \Sigma_2 \> \> , \> \> r \Def  \frac{1}{2} \left (\sigma_1 - \sigma_2 \right ) \> \> ,
\> \> s \Def  \frac{1}{2} \left (\sigma_1 + \sigma_2 \right ) \> \ge 0 \> .
\label{def S,r,s}
\ee
Then one has
\be
a_{12} \E \frac{1}{2} \, \Bigl [ \, |S+s| + |S-s| - |S+r| - |S-r| \, \Bigr ] 
\E \left \{ \begin{array}{r@{\quad  {\rm for} \quad} l}
                   s - |r| & |S| \le |r| \\
                   s - |S| & |r| \le |S| \le s \\
                   0       & |S| \ge s \> . \end{array} \right.
\label{a12}
\ee
\vspace{0.2cm}

\begin{figure}[hbt]
\vspace*{-1cm}
\bce
\mbox{\epsfxsize=120mm \epsffile{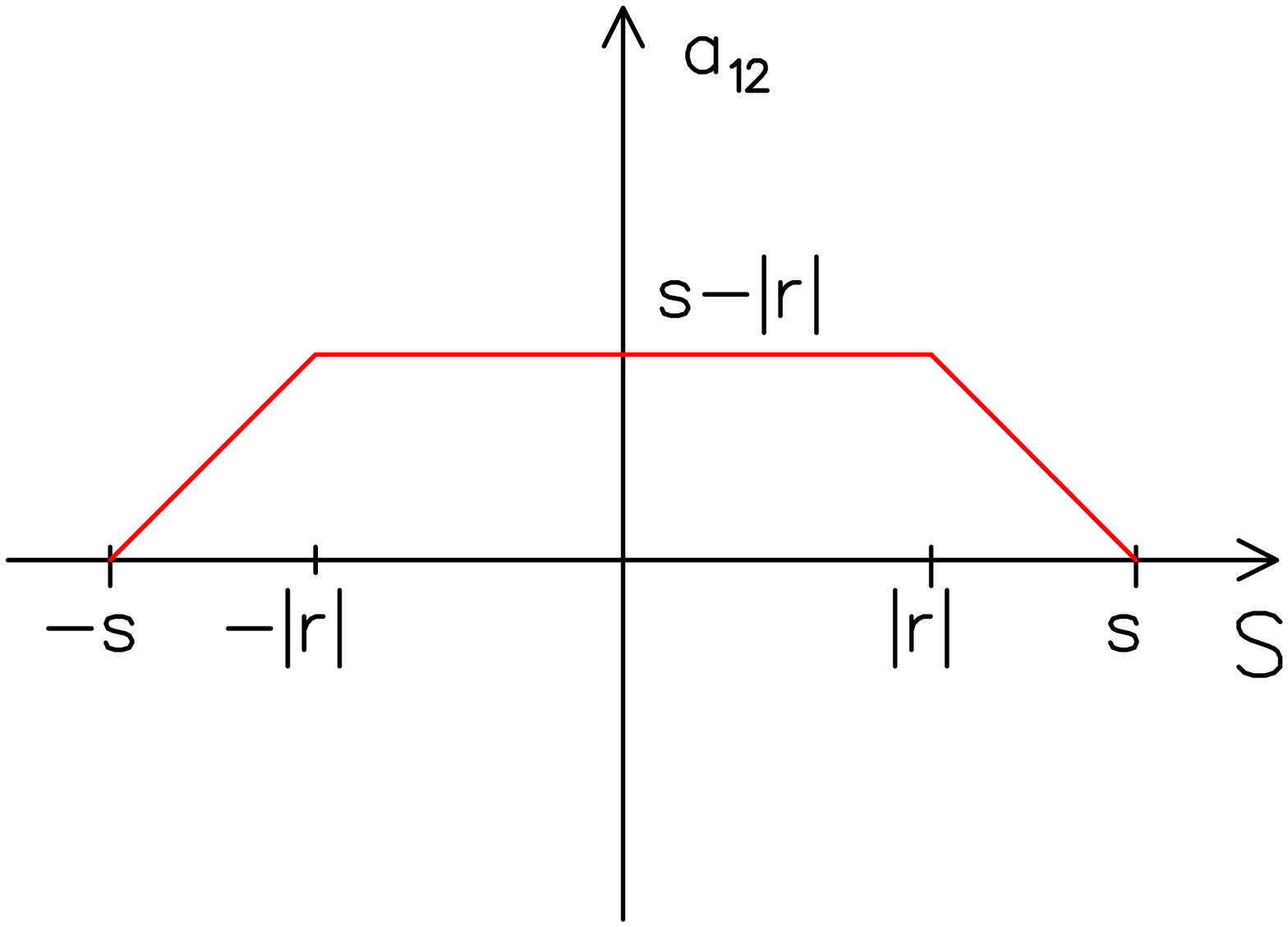}}
\ece
\vspace*{-1.6cm}
\caption{(Color online) The nondiagonal matrix element $a_{12}$ of the matrix (A) as a function of the time variables 
defined in Eq. (\ref{def S,r,s}).}
\vspace*{1.2cm}
\label{fig: a12}
\end{figure}

Figure \ref{fig: a12}  shows that $a_{12}$ 
is indeed a nonanalytic function of the times as expected from the absolute 
values in Eq. (\ref{a_ij}). Note that it is even in $S, r, s$.
If we would split up the integration
region into subregions where the time differences have definite sign we would get rid of that
complication at the price of considering many different contributions. This is exactly what happens
in the diagrammatic approach and is the source of the proliferation of diagrams
in high-order perturbation theory.

\section{How to make the BP approach numerically feasible}
\label{sec: tricks}

Equation (\ref{en limit}) together with Eqs. (\ref{cum recurs}) and (\ref{m_n}) specify how to calculate
the $n$th-order coefficient $e_n$ for the perturbative expansion of the polaron ground-state energy.
Taken at face value one needs to evaluate a $3n$-dimensional integral at large (asymptotic) 
values of the 
inverse temperature $\beta$. While this seems doable in principle, it is clear that in practice 
precise values of $e_n$ or the numerical feasibility of the whole approach need further 
improvements and refinements. As these practical questions have not been adressed at all
in Bogoliubov and Plechko's paper \cite{BoPl} we will describe several steps crucial for success.

\subsection{Additional integrations}
It is obvious that any reduction in the dimensionality of the integral to be evaluated numerically 
will be of great help. As explained above the integrations over the times can only be performed
by splitting the integration regions in many subregions leading to the time-honored
diagrammatic approach.
However, the dependence on the auxiliary variables $u_i$ is simple and analytic and 
therefore it is possible
to perform some of the integrations over them by expanding the $n \times n$ determinant 
$ \> {\rm det}_n A \> $ into cofactors \cite{Greub}.
For example, the dependence on $u_n$ is simply obtained by expanding with respect to 
the $n$th row (or column)
\be
{\rm det}_n \, A \E u_n \, A_n + {\rm det}_n \, A(u_n = 0) \> ,
\ee
where  $ \> A_n \> $ denotes the determinant of the matrix which is obtained from
$(A)$ by removing the  $n$th row and the $n$th column, i. e., it is a special  
$(n-1) \times (n-1)$ determinant known as principal minor \cite{fn_4}.
Therefore the integration over $u_n$ in Eq. (\ref{m_n}) can be easily performed:
\be
{\cal D}_n(1,2,\dots,\underline{n}) \Def 
\int_0^{\infty} du_n \> {\rm det}_n^{-3/2} A (1,2,\ldots,n)
\E \frac{2}{A_n \sqrt{{\rm det}_n \, A(u_n=0)}} \> .
\ee
Here we use the short-hand notation $ i \Def (t_i,t_i',u_i) $ and 
the integration over $u_n$ is indicated by underlining the $n$th argument. 
The dependence on $u_{n-1}$ is obtained similarly:
\bea
A_n \EA u_{n-1} \, A_{n-1,n} + A_n \left (u_{n-1}=0 \right ) \> , \\
{\rm det}_n A(u_n=0) \EA u_{n-1} \, A_{n-1}(u_n=0) + {\rm det}_n A \left ( u_{n-1}=u_n=0 \right ) \> .
\eea
Here $ \> A_{n-1,n} \> $ denotes the determinant (principal minor) of the matrix which is obtained from 
$(A)$ by removing both the  $(n-1)$th and the $n$th row and column. 
The subsequent integration over $u_{n-1}$ is therefore still an elementary one 
($u_{n-1} = u_n = 0$ is understood in all determinants from now on)
\bea
{\cal D}_n(1,2,\ldots,\underline{n-1},\underline{n}) &:=& \int_0^{\infty}du_{n-1} \int_0^{\infty} du_n 
\> {\rm det}_n^{-3/2} A (1,2,\ldots,n) \non
\EA \int_0^{\infty}du_{n-1}\> \frac{2}{ u_{n-1} \, A_{n-1,n} + A_n}
\>  \frac{1}{\sqrt{ u_{n-1} A_{n-1} + {\rm det}_n \, A}}
\eea
but depends on the sign of the combination $ A_n A_{n-1}  - A_{n-1,n} \, {\rm det}_n A $.
This is fixed since all the coefficients in the integrand are 
principal minors  of the positive semidefinite matrix $(A)$ which not only are non-negative
themselves but also obey the Hadamard-Fischer inequality [Ref. \cite{HoJo}, Eq. 7.8.9]
\be
A_{n-1} \, A_n \> \ge  \> A_{n-1,n} \, A
\label{Had-Fish}
\ee
($A \equiv {\rm det}_n \, A$ ).
Therefore the integration over $u_{n-1}$ gives [see, e.g., Ref. \cite{Dwight}, Eq. 192.11]
\be
{\cal D}_n(1,2,\ldots,\underline{n-1},\underline{n})
\E \frac{4}{ \sqrt{A_{n-1,n} \, A_{n-1} \, A_n}} \>  \frac{\arcsin \sqrt{x_{HF}}}{\sqrt{x_{HF}}} \> ,
\label{u_n-1 int}
\ee
where
\be
 0 \> \le \> x_{HF} \Def 1 - \frac{A_{n-1,n} \, A}{A_{n-1} \, A_n } \>  \le \> 1
\label{x_HF^2}
\ee
is non-negative and does not exceed unity as needed for a proper argument of the $\arcsin$ function. 
\vspace{0.3cm}

Let us illustrate that for the case $ n = 2 $ where all principal minors can be evaluated easily.
With Eqs. (\ref{cum2}) and (\ref{m_n}) one then obtains
\bea
\lambda_2 \EA \frac{\alpha^2}{4 \pi} \int_0^{\beta} dt_1 \, dt_2 \int_0^{t_1}dt_1' \int_0^{t_2}dt_2' \> 
e^{-(t_1+t_2-t_1'-t_2')} \> \left [ \, {\cal D}_2(\underline{1},\underline{2}) - 
{\cal D}_1(\underline{1}) \,{\cal D}_1(\underline{2}) \, \right ] \non
\EA \frac{\alpha^2}{\pi} \int_0^{\beta} dt_1 \, dt_2 \int_0^{t_1}dt_1' \int_0^{t_2}dt_2' \> 
e^{-(t_1+t_2-t_1'-t_2')} \> \frac{1}{\sqrt{a_{11} a_{22}}} \, 
f_2 \left ( \frac{a_{12}}{\sqrt{a_{11} a_{22}}} \right ) \> ,
\label{lambda2}
\eea
where 
\be
f_2(x) \Def \frac{\arcsin(x)}{x} - 1 \> .
\ee
Thus the second cumulant (and therefore the second energy coefficient) would vanish without 
the nondiagonal matrix element $a_{12}$, i.e., the correlation between the times when the
two phonons have been emitted(absorbed).

\subsection{Extrapolation for $ \beta \to \infty$}
A crucial question for the feasibility of the BP approach is how the asymptotic 
limit $\beta = \infty$ is reached.
From Appendix \ref{app: m12} where the cases $ n = 1, 2$ are treated explicitly we expect
\be
\lambda_n(\beta) \To \beta \, e_n  + d_n + 
{\cal O} \left ( e^{-\beta}/\sqrt{\beta} \right ) 
\ee
so that from Eq. (\ref{en limit}) only a rather slow convergence to the asymptotic
value is expected:
\be
e_n \E \lim_{\beta \to \infty} \, \left [ \, e_n + \frac{d_n}{\beta} \, \right ] \> .
\ee
This can be greatly improved not by dividing $\lambda_n(\beta)$ by $\beta$ but 
by taking the {\it derivative} of $\lambda_n(\beta)$, i. e., considering
\be
e_n(\beta) \Def \frac{(-)^{n+1}}{ \alpha^n n !}\lim_{\beta \to \infty} \,
\frac{\partial \lambda_n (\beta)}{\partial \beta} 
\label{en by deriv}
\ee
which approaches the asymptotic value {\it exponentially} 
\be
e_n(\beta)  \> \stackrel{\beta \to \infty}{\longrightarrow} \>  \frac{\partial}{\partial \beta}
\, \Biggl [ \, \beta   e_n + d_n + {\cal O} \left ( e^{-\beta}/\sqrt{\beta} \right ) \, \Biggr ]
\E e_n +  {\cal O} \left ( e^{-\beta}/\sqrt{\beta} \right )
\ee
 -- at least in the analytical examples given in Appendix \ref{app: m12} for  $ n = 1, 2$.

\noindent
We therefore will assume that for large enough $\beta$
\be
e_n(\beta) \To e_n + \frac{a_n}{\sqrt{\beta}} \, e^{-\beta}
\label{en asy}
\ee
for all values of $n$ in the following. Alternatively, the behavior
\be 
e_n(\beta) \> \stackrel{\beta \to \infty}{\longrightarrow} \>  e_n + \frac{a_n}{\beta^{\nu_n}} 
\, e^{-\beta}
\label{en asy gen}
\ee
will be fitted to the numerical data if they are precise enough to determine also the
power $\nu_n$.

Moreover, evaluating the differentiation with respect to $\beta$ also lowers the dimension of the 
integral which has to be evaluated numerically because the variable $\beta$ enters as upper limits
of the multidimensional integral (\ref{m_n 2}). Writing the corresponding cumulant as
\be
\lambda_n \deF (-)^n \frac{\alpha^n}{(4 \pi)^{n/2}} \, \prod_{j=1}^n \left ( \int_0^{\beta}
d\sigma_j \int_{\sigma_j/2}^{\beta - \sigma_j/2} d\Sigma_j \right ) \> 
F_n \left (\sigma_1,\Sigma_1;\sigma_2,\Sigma_2;\ldots;\sigma_n,\Sigma_n \right )
\ee
we find no contribution by differentiating the upper limit of the $\sigma_j$ integration 
since the range of $\Sigma_j$ then vanishes. Thus
\be
\frac{\partial \lambda_n}{\partial \beta} \E  (-)^n \frac{\alpha^n}{(4 \pi)^{n/2}} \,
\prod_{j=1}^n \left ( \int_0^{\beta} d\sigma_j \right ) \,  \sum_i^n \, \prod_{k \ne i} 
\left ( \int_{\sigma_k/2}^{\beta - \sigma_k/2}d\Sigma_k \right )
\> F_n \left (\sigma_1,\Sigma_1;\sigma_2,\Sigma_2;\ldots;\sigma_n,\Sigma_n \right )
\Bigr|_{\Sigma_i = \beta - \sigma_i/2}
\label{diff lambdan}
\ee
For example, for $n = 2 $ we have
\be
\frac{\partial \lambda_2}{\partial \beta} \E  \frac{ \alpha^2}{ \pi}
\int_0^{\beta} d\sigma_1 d\sigma_2 \,   
\frac{1}{\sqrt{\sigma_1 \sigma_2}} \,
 e^{-(\sigma_1 + \sigma_2)} \, \left [ \, \int_{\sigma_2/2}^{\beta - \sigma_2/2} d \Sigma_2
\>  f_2 \left ( \frac{a_{12}}{\sqrt{\sigma_1 \sigma_2}} \right )_{\Sigma_1 = \beta
- \sigma_1/2} + ( 1 \leftrightarrow 2) \, \right ] \> .
\label{diff lambda2}
\ee

\subsection{Symmetrization}

We may exchange simultaneously 
\be
\sigma_j, \Sigma_j  \> \leftrightarrow \> \sigma_k, \Sigma_k  \> \> , j \neq k = 1, \ldots , n 
\ee
in the integrand of Eq. (\ref{diff lambdan}).
There are $n!$ ways of doing that and thus 
\be
\frac{\partial \lambda_n}{\partial \beta} \E  \frac{(-\alpha)^n}{(4 \pi)^{n/2}} \,
\prod_{j=1}^n \left ( \int_0^{\beta}
d\sigma_j \right ) \,  \sum_i^n \, \prod_{k \ne i} 
\left ( \int_{\sigma_k/2}^{\beta - \sigma_k/2}d\Sigma_k \right )
\> \frac{1}{n!} \, \underbrace{\sum_{{\rm permut.}} \, 
F_n \left(\{ \sigma_j,\Sigma_k \} \right )}_{\deF F_n^{\rm symm} \left(\{ \sigma_j,\Sigma_k \} \right )}
\Biggr |_{\Sigma_i = \beta - \sigma_i/2}
\ee
and the domain of integration can be reduced \cite{fn_5}:
\be
\frac{\partial \lambda_n}{\partial \beta} \E  \frac{(-\alpha)^n}{(4 \pi)^{n/2}} \,  
\int_0^{\beta} d\sigma_1 \! \int_0^{\sigma_1} d\sigma_2 \ldots \! \int_0^{\sigma_{n-1}} d\sigma_n \, 
 \sum_i^n \, \prod_{k \ne i} \left ( \int_{\sigma_k/2}^{\beta - \sigma_k/2}d\Sigma_k \right ) \, 
F_n^{\rm symm} \left(\{ \sigma_j,\Sigma_k \} \right )\Biggr |_{\Sigma_i = \beta - \sigma_i/2} \, .
\label{diff lambdan symm}
\ee
Again taking $n = 2 $ as simple example we find from Eq. (\ref{diff lambda2}) that
$ F_2^{\rm symm} = 2 F_2 $ as the integrand is already completely symmetric. Hence
\be
\frac{\partial \lambda_2}{\partial \beta} \E \frac{2 \alpha^2}{\pi} \int_0^{\beta} d\sigma_1 \, 
\int_0^{\sigma_1} d\sigma_2  \> 
\frac{\exp(-\sigma_1-\sigma_2)}{\sqrt{\sigma_1 \sigma_2}} \, \left [ \, 
\int_{\sigma_1/2}^{\beta - \sigma_1/2}d\Sigma_1  \, f_2 \left ( 
\frac{a_{12}}{\sqrt{\sigma_1 \sigma_2}} \right )\Biggr |_{\Sigma_1 = \beta - \sigma_1/2} 
+ ( 1 \leftrightarrow 2) \, \right ]
\label{diff lambda2 symm}
\ee
where we have used $a_{ii} = \sigma_i $. Further evaluation of Eq. (\ref{diff lambda2 symm})
is presented in Appendix \ref{app: m12}.
For $ n > 2 $ we have to perform the symmetrization explicitly as the integrations over 
$u_n, u_{n-1}$ lead to a nonsymmetric integrand.

\subsection{Mapping}

Finally for Monte Carlo integration
we need a mapping to bring all integration variables into the hypercube $[0,1]$. After some
experimentation we have chosen
\be
u_i \E \sigma_i \left ( \frac{1}{\xi_i^2} - 1 \right ) \> , \> i = 1, 2, \ldots , (n-2) 
\ee
and 
\bea
\sigma_1 \EA \beta s_1^2 \> , \>  \> \sigma_i \E \sigma_{i-1} \, s_i^2 \> , \> \> i = 2, \ldots , n \> ,
\label{mapping sigma}\\
\Sigma_i \EA \left ( \beta - \sigma_i \right ) \, S_i + \frac{1}{2} \sigma_i
\label{mapping Bigsigma}
\eea
as transformation of the remaining variables. Here all $ \xi_i, s_i, S_i \in [0,1] $.
Equation (\ref{mapping sigma}) removes possible  
square-root singularities which are seen in the examples for $ n = 1, 2 $ in Appendix \ref{app: m12} -- 
these are integrable analytically but would pose severe problems for numerical integration. More refined
mappings of the relative times (for example, to include the exponential suppression) have been
tried but did not result in significant improvements.


\section{Numerical results}
\label{sec: num}

\subsection{A test: ${\bf e_3}$}
We have tested our approach by determining the third order coeffcient $e_3$ which has been calculated 
by Smondyrev \cite{Smon} with later improvements in accuracy \cite{SeSm}. Table \ref{tab: e3} lists 
the values of $e_3(\beta)$ obtained by Monte Carlo integration using the classic VEGAS program 
\cite{Vegas} with $n_{MC} = 4.9 \times 10^8$ function calls per iterations. We have used 100 iterations 
for each $\beta$ value. Thus the total number of function calls was
\be
n_{tot}^{(3)} \E n_{MC} \,  \, n_{it} \E  4.9 \times 10^{10} \> .
\ee
\begin{figure}[hbt]
\bce
\mbox{\epsfxsize=90mm \epsffile{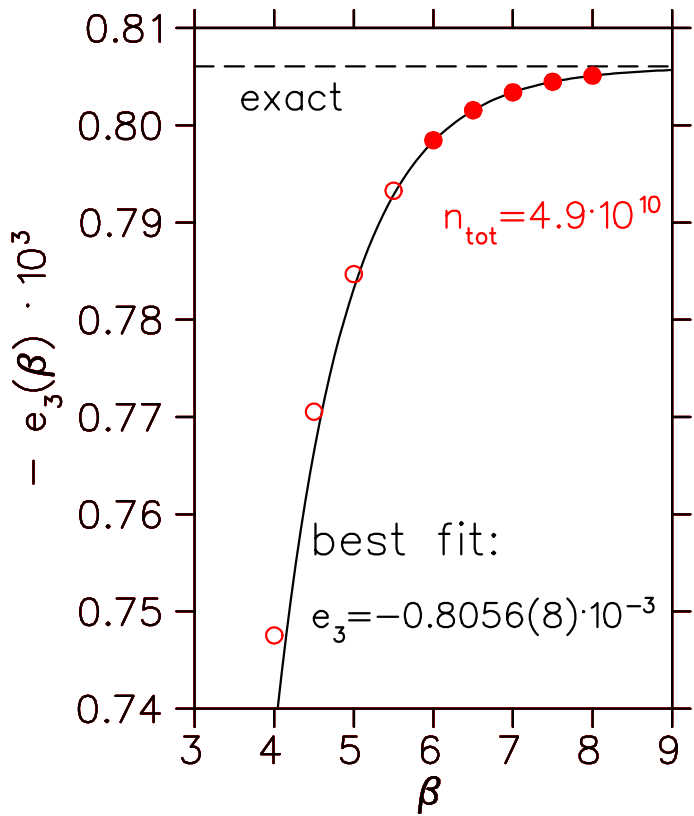}}
\ece
\caption{(Color online) Monte Carlo results for the derivative of the third cumulant
as a function of the Euclidean time (inverse temperature) $\beta$. 
The total number of function calls is denoted by $n_{\rm tot}$
and the full (open) circles are the points used (not used) in the fit (see Table \ref{tab: fit e3}.}
\label{fig: cum3}
\vspace{0.6cm}
\end{figure}

Figure \ref{fig: cum3} shows that $e_3(\beta)$ monotonically approaches Smondyrev's value with
increasing $\beta$. The sheer fact that $e_3(\beta)$ converges to a constant value at large
$\beta$ is a good signal: individual moments $m_n$ would behave 
as $\beta^n$ for large values of $\beta$ but the construction of the 
cumulants takes away all these $\beta$ powers except the linear one which contains
the information about the ground-state energy.
\begin{table}[hbt]
\caption{Third order energy coefficient $e_3(\beta)$ from the
derivative of the the third cumulant
as a function of the inverse temperature $\beta$. The numerical results were
obtained with the Monte Carlo routine VEGAS for evaluating the full six-dimensional
integral. Numbers in parenthesis are the estimated errors in units of the last digit. The last
column gives the $\chi^2$ per degree of freedom ($N_{DF}$) monitored during the iterations.
This should be close to one if the iterations are consistent with each other.}

\begin{center}
\begin{tabular}{ccc} \hline\hline
                      &                           &              \\

\hspace{0.5cm} $\beta$ \hspace{0.5cm} & $ -e_3(\beta) \times 10^3$  &  \quad \quad $\chi^2/N_{DF}$
\hspace{0.5cm} \\
                      &                           &               \\ \hline
                      &                           &               \\

4.0 \quad             &  \quad   0.7474 ( 5)    &  \quad  0.969        \\
4.5 \quad             &  \quad   0.7704 ( 7)    &  \quad  0.876         \\
5.0 \quad             &  \quad   0.7846 ( 8)    &  \quad  0.836        \\
5.5 \quad             &  \quad   0.7934 (10)    &  \quad  0.837         \\
6.0 \quad             &  \quad   0.7987 (11)    &  \quad  0.821        \\
6.5 \quad             &  \quad   0.8017 (13)    &  \quad  0.792      \\
7.0 \quad             &  \quad   0.8033 (15)    &  \quad  0.768        \\
7.5 \quad             &  \quad   0.8039 (17)    &  \quad  0.775          \\
8.0 \quad             &  \quad   0.8041 (19)    &  \quad  0.772         \\
                      &                           &                       \\ \hline\hline

\end{tabular}
\end{center}
\label{tab: e3}
\end{table}

\begin{table}[hbt]
\caption{Extrapolation of $e_3(\beta)$ to $\beta = \infty$ 
using the data from Table \ref{tab: e3}, the fitting range $  \beta \in [\beta_{min},\beta_{max}]$ and 
the fixed power $\nu_3 = 0.5$ in the ansatz (\ref{en asy}). The last column gives the
$\chi^2/N_{DF}$ of the two-parameter fit where $N_{DF}$ = number of data points - 2.
}
\begin{center}
\begin{tabular}{llcc} \hline\hline
               &                 &                    &  \\

\quad  $\beta_{\rm min}$  &  $\beta_{\rm max}$  & $ -e_3 \times 10^3$  &\quad \quad 
$\chi^2/N_{DF}$\hspace{0.5cm}\\
               &                 &                    &  \\ \hline
               &                 &                    & \\

\quad 4.0 \quad      & 8.0 \quad       &  \quad   0.8043 \, (6)    &  \quad  0.989    \\
\quad 4.5 \quad      & 8.0 \quad       &  \quad   0.8052 \, (7)    &  \quad  0.138         \\
\quad 5.0 \quad      & 8.0 \quad       &  \quad   0.8056 \, (8)    &  \quad  0.048        \\
\quad 5.5 \quad      & 8.0 \quad       &  \quad   0.8055 (10)      &  \quad  0.058        \\

               &                 &                        &                       \\ \hline\hline

\end{tabular}
\end{center}

\label{tab: fit e3}
\end{table}

\clearpage

We have fitted these data with the ansatz (\ref{en asy}) which, of course, only holds for 
asymptotic values of $\beta$. Therefore the lower limit $\beta_{\rm min}$ 
of the fit range $[\beta_{\rm min},\beta_{\rm max}]$ was successively raised until the 
$\chi^2/N_{DF}$ of the fit reached a minimum. This is displayed in Table \ref{tab: fit e3}.
If $\beta_{\rm min}$ is too close to $\beta_{max}$ the degrees of freedom decrease which should
cause the $\chi^2/N_{DF}$ to increase in turn \cite{fn_6}.
This fitting strategy yielded
\be
e_3 \E   - 0.8056 (8)   \times 10^{-3} \> .
\label{e3 n=-0.5}
\ee
If we allow the more general ansatz (\ref{en asy gen}) we obtain as best fit
\be
e_3 \E - 0.8055 (6)  \times 10^{-3} 
\label{e3 n free}
\ee 
and
\be
\nu_3 = 0.55 (3) \> .
\ee
The above error estimates may be a little bit optimistic since we have taken the VEGAS
errors at face value. In addition, the power $\nu_3$ and the parameter $a_3$
in the fit function (\ref{en asy gen}) turn out to be highly correlated. Nevertheless the 
$\exp(-\beta)/\sqrt{\beta}$ behavior also seems to hold for higher cumulants and the extrapolated 
result is in good agreement with Smondyrev's 
analytical result (\ref{e3}). The main message of this test therefore is 
that (our implementation of) the BP method is working and able to give accurate values for  
the perturbative expansion of the ground-state energy of a polaron.

\subsection{A new coefficient: ${\bf e_4}$}
\begin{figure}[hbt]
\bce
\mbox{\epsfxsize=90mm \epsffile{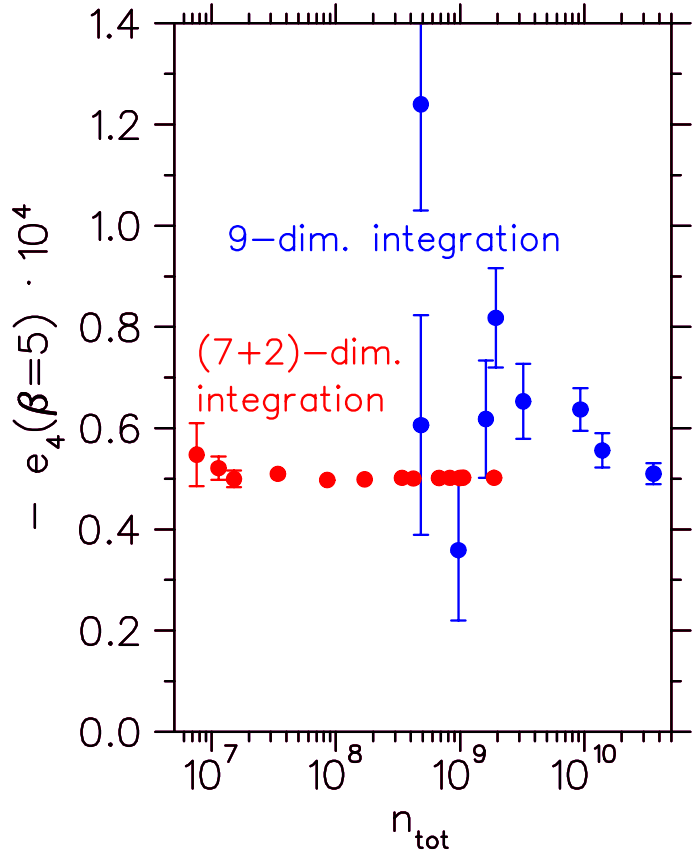}}
\ece
\caption{(Color online) Convergence of the fourth order coefficient $e_4(\beta=5)$ for a fixed value 
of the inverse temperature $\beta$ as a function of the total numer of function calls $n_{\rm tot}$. 
Square (blue) points denote
the case where the full nine-dimensional integral was evaluated by the Monte Carlo routine VEGAS, 
(red) circles show the result if two of the integrations are done by the deterministic 
tanh-sinh-quadrature rule and the
rest stochastically.
}
\vspace{0.8cm}
\label{fig: converg4}
\end{figure}

When applying the previous approach to the calculation of the first unknown coefficient $e_4$ an 
unpleasant outcome is found: as seen in Fig. \ref{fig: converg4} for a  fixed value of $\beta = 5 $ 
the convergence with the
number of function calls is very slow. Since the cancellations in the integrand are more severe
for the large $\beta$ which is needed for determining $e_4$ only a very rough determination of this
coefficient was possible in acceptable CPU time.

Fortunately a solution was found by performing the remaining integrations over 
$u_i \> , i = 1,2 $ by a {\it deterministic}
integration routine. While such an option is not available for the time integrations for which the 
integrand is nondifferentiable (see Fig. \ref{fig: a12}) it is possible for the integration over the 
auxiliary variables $u_i$ where the dependence is an analytic one [see Eqs. (\ref{m_n}, \ref{(A)})].

\begin{figure}[hbt]
\bce
\mbox{\epsfxsize=90mm \epsffile{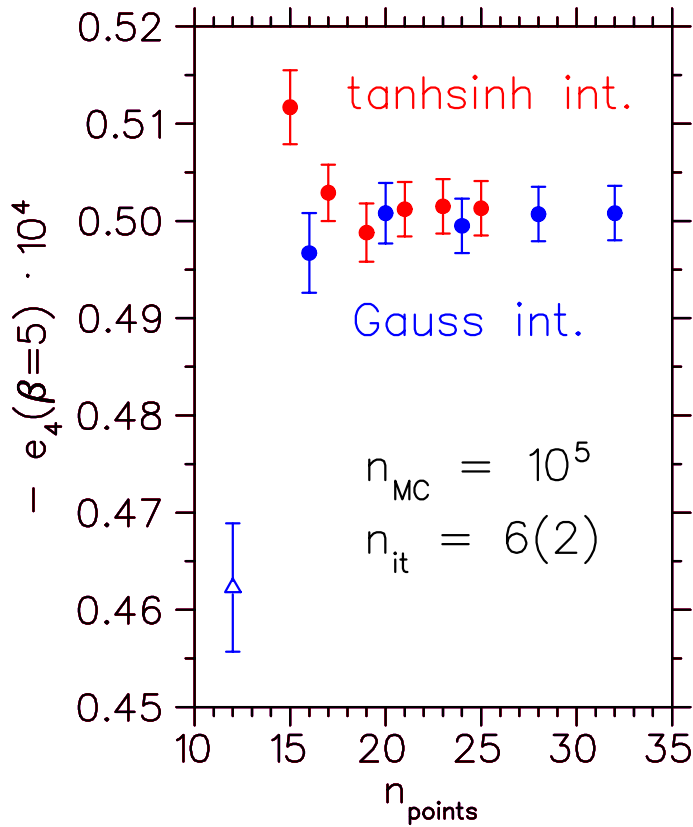}}
\ece
\caption{(Color online) Comparison of deterministic integration routines for $e_4(\beta = 5)$ as 
a function of the
number of integration points. The number of Monte Carlo calls ($n_{MC}$) and 
iterations ($n_{it}$) is kept fixed. An open symbol indicates a Monte Carlo result with
inconsistent iterations. 
}
\vspace{0.6cm}
\label{fig: comp4_tanh_gauss}
\end{figure}

We have used the powerful tanh-sinh integration procedure \cite{TaMo} 
which -- after a judicious transformation
of variables -- is nothing else than the trapezoidal approximation to the transformed
integral
\be
\int_a^b dx \> f(x) \> \approx \>  h \, \frac{b-a}{2} \sum_{k=-k_{max}}^{k_{max}} \, w_k
                  \, f \left ( \frac{b+a}{2} + \frac{b-a}{2} x_k \right )
\label{tanhsinh a,b}
\ee
with precalculated abscissae $x_k$ and weights $w_k$.
Since this quadrature rule seems not to be very well known (see, however, 
Ref. \cite{tanhsinh}) Appendix \ref{app: tanh} gives a short account of its basic features 
together with details 
of our implementation. Having in mind an application to our
multidimensional case the convergence rate with the number of function calls
\be
n_t \E 2 \, k_{max} + 1
\ee
is of paramount interest. In the one-dimensional case 
the error may decrease as fast as $ \exp( - c \, n_t/\ln n_t) $ \cite{optimal,error}
depending on the analyticity domain of the transformed function $f(x)$.
However, without any knowledge about that and in a multidimensional
application, such an error estimate is of no help and we have to test the convergence of the 
quadrature rule with increasing $n_t$. The outcome is also shown in Fig. \ref{fig: converg4} 
as function of
\be
n_{tot}^{(4)} \E n_t^2 \,   n_{MC} \,  n_{it}
\ee
and demonstrates an improvement by two orders of magnitude compared to the previous approach
which fully evaluated the nine-dimensional integral by stochastic methods. Figure 
\ref{fig: comp4_tanh_gauss} shows a comparison
with Gaussian integration which also gives fairly good results.

\vspace{1.5cm}

This improvement now allows a much more precise determination of the coefficient $e_4 $ 
(and, of course, also of the third order coefficient \cite{fn_7}).
Table~\ref{tab: e4}  contains the data for $e_4(\beta)$ from $ \beta = 4 $ to $ \beta = 8 $ 
each with 12 iterations; the first 2 iterations were used
for establishing the optimal grid while the following 10 were utilized for the statistics [denoted by
$ n_{it} = 12(2)$ in the following]. In addition to the classic VEGAS program (as in the previous 
test for $e_3$) we also have used the VEGAS program from the CUBA library 
\cite{Cuba} which employs Sobol quasirandom numbers. This allowed to extend the range
of inverse temperatures up to $\beta = 10$. 
Typical run times were about 1 day on a 2.4 GHz PC.
It is seen that for all $\beta$ there is agreement between the two data sets within the error bars. 
Despite larger statistics and higher accuracy in the deterministic integration the
VEGAS (Cuba) routine returns larger errors which reflects our experience 
that the VEGAS (classic) error estimate often is too optimistic. This is also corroborated by 
the observation that 
at various $\beta$-values the VEGAS (classic) results have an unacceptable large 
$\chi^2/N_{DF}$ 
indicating inconsistencies between
different iterations within the given error bars.

\begin{figure}[hbt]
\bce
\mbox{\epsfxsize=90mm \epsffile{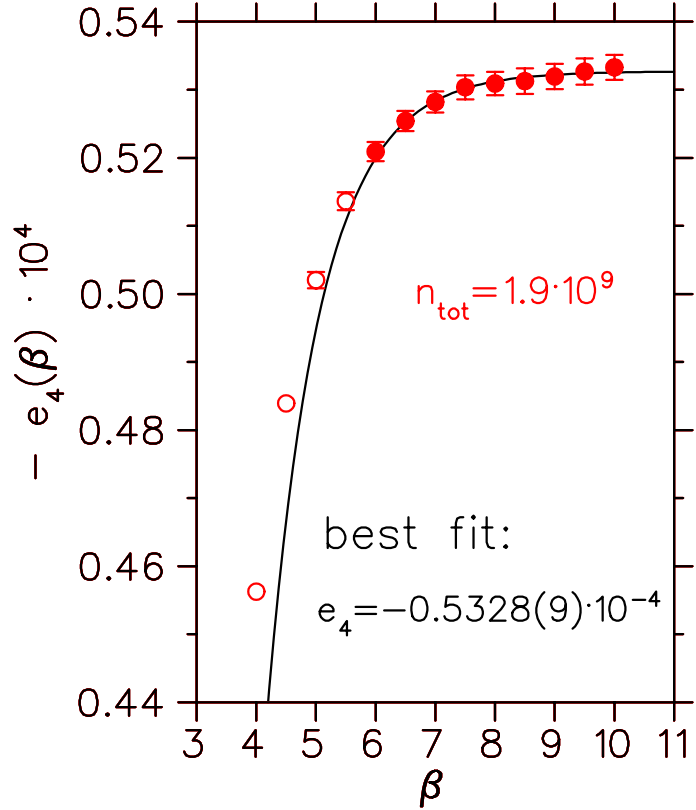}}
\ece
\caption{(Color online) Same as in Fig. \ref{fig: cum3} but for the derivative of the fourth cumulant.
The plotted data points are the VEGAS (Cuba) results from Table \ref{tab: e4}.
}
\label{fig: cum4}
\vspace{0.6cm}
\end{figure}

\begin{table}[hbt]
\caption{Same as in Table \ref{tab: e3} but for the fourth-order term $e_4(\beta)$.
The numerical results were obtained by a combination of deterministic and stochastic integration 
of the nine-dimensional integral (see text). Two different versions of the VEGAS program have been used:
the classic one with pseudo-random numbers 
and the CUBA version with Sobol quasirandom numbers. The number of points in the
deterministic tanh-sinh integration is denoted by $n_t$. In the VEGAS (classic) evaluation
$ n_{it} = 12(2) $ iterations were used at each $\beta$ value.
Data marked by an asterisk  have an unacceptable
$\chi^2/N_{DF}$ (underlined) indicating that the iterations do not lead to a consistent error estimate.
The last column gives the probability $p$ that the error estimate for the VEGAS (Cuba) results 
is not reliable ($ p < 0.95$ is considered to be safe).
}
\begin{center}
\begin{tabular}{cclcl} \hline\hline
                      &                            &                 &                        & \\
                      &    \quad VEGAS (classic):  &  $n_{MC} = 4.7 \times 10^5$ &   
\quad VEGAS (Cuba): & $n_{MC} = 3 \times 10^6 $      \\ 
                      &                            &  $n_t = 23$      &              & $n_t = 25$ \\
                      &                           &                 &                        &   \\
\quad  $\beta$ \quad  & $-e_4(\beta) \times 10^4$  &  \quad  $\chi^2/N_{DF}$  & 
$-e_4(\beta) \times 10^4$ &  \quad \quad    $p$ \\
                      &                           &                 &                        & \\ \hline
                      &                           &                 &                        &   \\
 \quad  4.0 \quad     &  \quad   0.4549 ( 6)    &  \quad  0.637  &  \quad 0.4563 (10) &  \quad  0.164  \\
 \quad  4.5 \quad     &  \quad   0.4828 ( 7)    &  \quad  0.995  &  \quad 0.4839 (11) &  \quad  0.157  \\
 \quad  5.0 \quad     &  \quad \,    0.5013 ( 8)  $^{\ast}$   &  \quad  \underline{1.404}  & 
\quad 0.5020 (12) &  
\quad  0.170   \\
 \quad  5.5 \quad     &  \quad   0.5129 ( 8)    &  \quad  1.087  &  \quad 0.5136 (13) &  \quad  0.406  \\
 \quad  6.0 \quad     &  \quad\,    0.5193 ( 9)  $^{\ast}$   &  \quad  \underline{1.739}  &  
\quad 0.5209 (14) &  
\quad  0.413   \\
 \quad  6.5 \quad     &  \quad \,   0.5239 ( 9)  $^{\ast}$   &  \quad  \underline{1.488}  &  
\quad 0.5254 (15) &  
\quad  0.480   \\
 \quad  7.0 \quad     &  \quad   0.5271 (10)    &  \quad  0.977  &  \quad 0.5287 (16) &  \quad  0.534  \\
 \quad  7.5 \quad     &  \quad \,   0.5293 (10) $^{\ast}$    &  \quad  \underline{1.830}  &  
\quad 0.5304 (18) &  
\quad  0.588   \\
 \quad  8.0 \quad     &  \quad \,   0.5309 (11)  $^{\ast}$   &  \quad  \underline{1.520}  &  
\quad 0.5309 (17) & 
\quad   0.646   \\
 \quad  8.5 \quad     &                         &                &  \quad 0.5313 (19) & \quad   0.387  \\
 \quad  9.0 \quad     &                         &                &  \quad 0.5320 (19) &  \quad  0.355  \\
 \quad  9.5 \quad     &                         &                &  \quad 0.5327 (20) &  \quad  0.483  \\
 \quad  10.0 \quad    &                         &                &  \quad 0.5333 (19) & \quad   0.553  \\
                      &                         &                &                      &\\ \hline\hline

\end{tabular}
\end{center}

\label{tab: e4}
\vspace{0.4cm}
\end{table}
\clearpage
But also for the VEGAS (Cuba) results the probability that the error is unreliable
increases with the value of $\beta$. This just reflects the fact that the cancellations inside
the integrand are becoming more and more challenging at high $\beta$. 
Fitting the VEGAS (Cuba) data with the asymptotic ansatz (\ref{en asy})
yields
\be
e_4 \E - 0.5328 (9) \, \times \, 10^{-4} \> .
\label{e4 n=-0.5}
\ee
Data and best fit are shown in Fig. \ref{fig: cum4}. The more general ansatz (\ref{en asy gen}) 
leads to
\be
e_4 \E - 0.5330 (7) \, \times \, 10^{-4} 
\label{e4 n free}
\ee
with $ \nu_4 = 0.35 (7) $. We therefore take
\be
e_4 \E - 0.533 (1) \, \times \, 10^{-4} 
\label{e4 final}
\ee
as our final result.

\vspace{1cm}

\subsection{A further step: ${\bf e_5}$}

We have extended the BP approach to the calculation of the fifth-order coefficient $e_5(\beta)$.
Numerically this is much more challenging than the fourth-order calculation since these coefficients
drop by roughly one order of magnitude in each order. This has to be achieved by cancellation in a 
12-dimensional integral over a much more complicated integrand leading to much larger CPU times.

\begin{figure}[hbt]
\vspace{0.5cm}
\bce
\mbox{\epsfxsize=90mm \epsffile{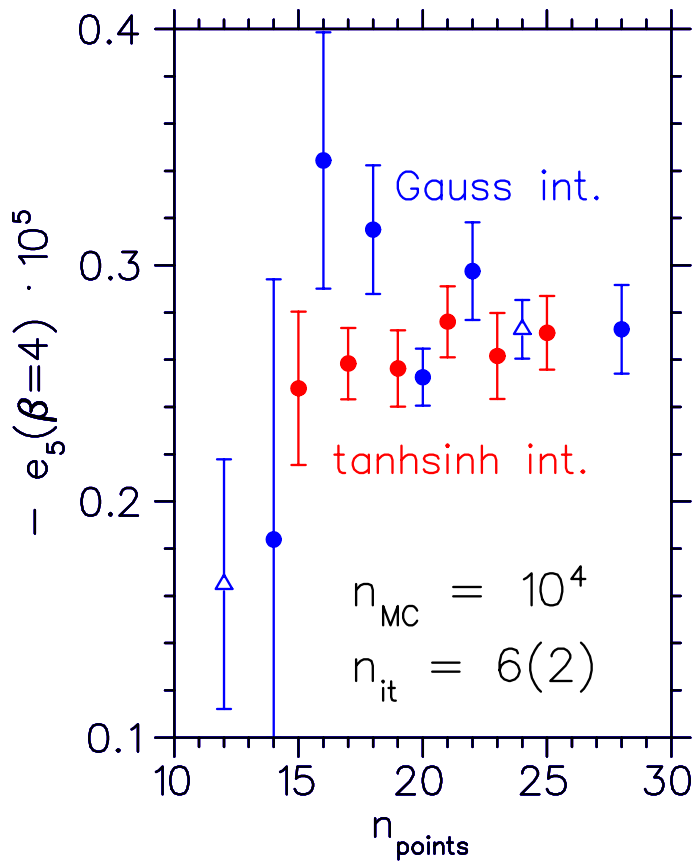}}
\ece
\caption{(Color online) Comparison of tanh-sinh and Gaussian integration for the derivative of the 
fifth cumulant at 
$\beta = 4$. Notation as in Fig. \ref{fig: comp4_tanh_gauss}.}
\label{fig: comp5_tanh_gauss}
\vspace{1cm}
\end{figure}

\noindent
Nevertheless the combination of deterministic integration and Monte Carlo
integration leads to reasonable results. Figure \ref{fig: comp5_tanh_gauss} shows a slight advantage of 
the tanh-sinh integration rule compared to Gaussian integration. Of course, due to the more severe 
cancellations in the 12-dimensional integrand higher accuracy, i.e., a larger number of 
deterministic integration points is needed. At the same time the number of Monte Carlo calls cannot be
as large as before to avoid excessive running times. 

Another numerical problem which already plagued the numerics for $ n = 4 $ (and to a much lesser extent
$n = 3$) became more severe in the present case: due to round-off errors the 
Hadamard-Fisher inequality (\ref{Had-Fish})
was not fulfilled exactly all the time: negative values down to
\be
x_{HF}^{\rm min} \E - 3.1 \times 10^{-9}
\ee
were recorded in double-precision arithmetic. Fortunately, this ``digit-deficiency error''
(see Appendix~B of Ref. \cite{Kinaccuracy}) does not
affect the outcome of the Monte Carlo runs:
checks have shown that
$e_5(\beta=4)$ comes out the same whether 
the negative argument is set to zero or the absolute value of $x_{HF}$ is taken. 
In addition, the use of quadruple precision gives a consistent result (within error bars) but
reduces the violation of the Hadamard-Fisher inequality considerably - 
at the price of a 20-fold longer running time.

\begin{figure}[hbt]
\vspace{1.5cm}
\bce
\mbox{\epsfxsize=90mm \epsffile{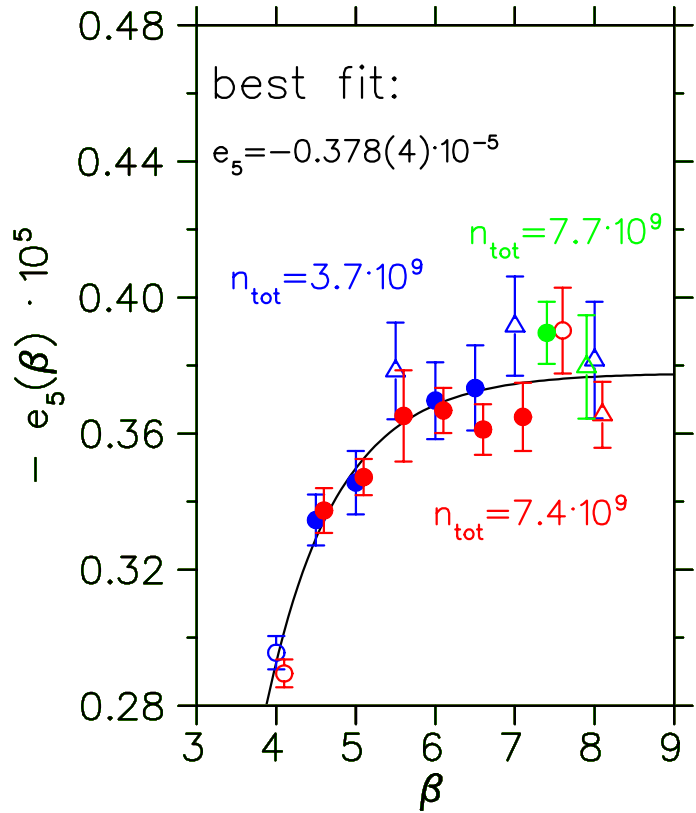}}
\ece
\caption{(Color online) Same as in Fig. \ref{fig: cum3} but for the derivative of the fifth cumulant.
Data points with open triangles are from statistically inconsistent Monte Carlo iterations 
(see Table \ref{tab: e5}) and are not used in the fit.}
\vspace{1cm}
\label{fig: cum5}
\end{figure}

\begin{table}[hbt]
\caption{Same as in Table \ref{tab: e4}  but for the fifth-order term $e_5(\beta)$. For all
deterministic numerical integrations  $n_t  = 25 $ integration points were used 
in the tanh-sinh integration routine. The Monte Carlo integrations were either done with the 
VEGAS (Cuba) program  $ \left ( n_{MC} = 1.5 \times 10^5 \right )$ or the classic VEGAS routine with
$n_{MC} = 7.9 \times 10^4 , n_{it} = 6(2)$ except for the data in boldface for 
which $n_{MC} =  9.8 \times 10^4 , n_{it} = 5(2)$. 
}
\vspace{0.5cm}
\begin{center}
\begin{tabular}{ccccc} \hline\hline
                &                            &                    &                     &   \\
                &   \quad VEGAS (classic)    &                    & \quad VEGAS (Cuba)  &  \\
\quad  \quad $\beta$ \quad \quad & $-e_5(\beta) \times 10^5$  & $\chi^2/N_{DF}$   &  
$-e_5(\beta) \times 10^5$ 
& \quad $p$ \hspace{0.3cm} \\
                &                            &                    &                     &   \\   \hline
                &                            &                    &                     &   \\
4.0 \quad       &  \quad   0.290 ( 4)        &   \quad 0.240      & \quad 0.295 (10)    &  0.369 \\
4.5 \quad       &  \quad   0.337 ( 7)        &   \quad 1.052      & \quad 0.317 (25)    &  0.722 \\
5.0 \quad       &  \quad   0.347 ( 6)        &   \quad 0.537      & \quad 0.349 (18)    &  0.353 \\
5.5 \quad       &  \quad   0.365 (14)        &   \quad 0.177      & \quad 0.330 (22)    &  0.365 \\
6.0 \quad       &  \quad   0.367 ( 7)        &   \quad 0.287      & \quad 0.327 (26)    &  0.657 \\
6.5 \quad       &  \quad   0.361 ( 8)        &   \quad 0.846      & \quad \, 0.370 (18)$^{\ast}$   &  
\underline{0.956} \\
7.0 \quad       &  \quad   0.365 (10)        &   \quad 0.984      & \quad 0.394 (30)    &  0.404 \\
7.5 \quad       &  \quad   \,   0.390 (13)   &   \quad 1.296      & \quad 0.390 (42) &  
0.329 \\
                &  \quad  \,  {\bf 0.390 ( 9)}    &  \quad \, {\bf 0.592}   &                     &   \\
8.0 \quad       &  \quad   \, \, 0.366 (10) $^{\ast}$ &  \quad \underline{2.514}  &  \quad 0.367 (35) & 
0.326  \\
                &  \quad  \, \,  {\bf 0.380 (15)}$^{\ast}$ &  \quad \, \underline{\bf 1.755}  &   &    \\   
                &                            &                    &                     &\\ \hline\hline

\end{tabular}
\end{center}

\vspace{0.5cm}
\label{tab: e5}
\end{table}
The data are collected in Table \ref{tab: e5} and show that at high $\beta$ 
it becomes more and more difficult
to get consistent numerical results. Typical run times for each $\beta$ value were about 1 month 
on a 3.0 GHz Xeon machine. With the
Intel ifort compiler some loops could be vectorized leading to a reduction in CPU time by 
more than a factor of 2.
If we exclude the data with $\chi^2/N_{DF} > 1.3 $  and $ p > 0.9 $ we obtain from a fit
with $ \nu_5 = 0.5 $ fixed 
\be
e_5 \E -0.378 (4) \times 10^{-5} \> .
\label{e5 n=-0.5}
\ee
This is shown in Fig. \ref{fig: cum5} together with the corresponding values of
\be
n_{tot}^{(5)} \E n_t^3 \,  n_{MC} \, n_{it} 
\ee
for the different data from Table \ref{tab: e5}. It is not possible to determine the exponent $\nu_5$
unambigously from the data which scatter too much. Taking a range of reasonable values for $\nu_5$ 
we end up with
\be 
e_5 \E -0.38 (2) \times 10^{-5} 
\label{e5 final}
\ee
as final result for the fifth order energy coefficient. It is obvious that the given error is
more an educated (and conservative) guess than a precise outcome of the fit.
\vspace{0.5cm}

\newpage
\section{Conclusion and outlook}
\label{sec: conclus}

We have shown that the Bogoliubov-Plechkov (BP) approach to calculate perturbative
coefficients without diagrams works for the polaron problem (a simple field theory of electrons 
and phonons) if it is combined with several simple but crucial ``tricks'' 
to enhance the numerical feasibility and convergence. There is no indication that higher
cumulants are ``unbounded from below'' as was reported in Ref. \cite{LYBKN} in a much simpler anharmonic
oscillator model \cite{fn_8}.
It is worthwhile to point out the 
advantages and disadvantages of the BP  approach
compared to the standard perturbative method.

While in the diagramatic approach a factorial increasing number of
individual (zero-temperature) diagrams adds up to the final result, much fewer terms (moments)
(see , e.g., Eqs. (\ref{cum4}, \ref{cum5})) must cancel inside the finite-temperature integral 
in the BP approach to obtain a result which is linear in $\beta$ so that the perturbative 
ground-state energy of the polaron can be determined. Of course, diagrams can be calculated
exactly at zero temperature whereas in the BP approach the extrapolation $\beta \to \infty$
must be performed numerically. We have demonstrated that by evaluating the derivative
of the various cumulants, an exponential convergence to the zero-temperature limit can be 
exploited. Two new perturbative coefficients $e_4$ and $e_5$ for the ground-state energy 
of a polaron have been obtained in this way and compared to results from Feynman's approximate
treatment.

\noindent
It should be emphasized that the BP approach says nothing about the convergence of the
perturbative series as it works in a fixed order. For the polaron case it is known
that the ground-state energy is an analytic function of the coupling constant \cite{GeLo}
but this is not necessary and systems where the perturbative expansion is known (or suspected)
not to converge could be treated as well.
Indeed, there
is some hope that the methods which in the present work 
have been applied successfully for a simple nonrelativistic field theory 
may also be suited for relativistic field theories such as QED and QCD if these are formulated in 
the worldline formalism. Renormalization of the occuring divergencies is the main new challenge
which is under investigation.

\vspace{6cm}

\noindent
{\bf Acknowledgement:} Many thanks to Michael Spira who supplied his version of
the classic VEGAS program and to Valery Markushin for help with compiler optimization which
led to a considerable speed up of the calculations. I am also indebted to Dr. Plechko 
who informed me about his previous work in Ref. \cite{BoPl} and made some valuable
remarks.

\newpage

\renewcommand{\thesection}{Appendix \Alph{section}:}
\setcounter{section}{0}

\renewcommand{\theequation}{\Alph{section}\arabic{equation}}

\vspace{0.5cm}

\section{Energy coefficients from a quadratic trial action}
\setcounter{equation}{0}
\label{app: best}
Here we briefly describe the results obtained with Feynman's variational
method and with the best quadratic approximation \cite{best polaron}.
Employing Jensen's inequality and working out the various path integral averages one finds
that the true ground-state energy is below the variational energy
\be
E_0 \> \le \> E_t \E \Omega + V \> ,
\ee
where
\be
\Omega \E \frac{3}{2 \pi} \, \int_0^{\infty} dE \> \left [ \, \ln A(E) + \frac{1}{A(E)} - 1 \, 
\right ] \> , \hspace{0.3cm} V \E - \frac{\alpha}{\sqrt{\pi}} \, \int_0^{\infty} d\sigma \> 
\frac{\exp(-\sigma)}{[\mu^2(\sigma)]^{1/2}} \> .
\ee
Here $A(E)$ is the ``profile function'' which is related to the retardation function by
\be
A(E) \E 1 + 8 \, \int_0^{\infty} d\sigma \> f(\sigma) \, \frac{\sin^2(E \sigma/2)}{E^2}
\label{A from f}
\ee
and $\mu^2(\sigma)$ the ``pseudotime'' \cite{fn_9}
given by
\be
\mu^2(\sigma) \E \frac{4}{\pi} \, \int_0^{\infty} dE \> \frac{1}{A(E)} \, 
\frac{\sin^2(E \sigma/2)}{E^2} \> .
\label{amu from A}
\ee

\noindent
In Feynman's original work the retardation function is parametrized as 
\be
f_F(\sigma) \E C \, e^{-w \, \sigma}
\label{f Feyn}
\ee
which has the advantage that profile function, pseudotime and the kinetic term can be calculated 
analytically:
\be
A_F(E) \E \frac{v^2 + E^2}{w^2 + E^2} \> , \> \> \> 
\mu^2_F(\sigma) \E \frac{w^2}{v^2} \sigma + \frac{v^2-w^2}{v^3} \, \left ( 1 - e^{-v \sigma} \right )
\> , \>  \>  \> \Omega_F \E \frac{3}{4} \, \frac{(v-w)^2}{v} \> .
\ee
Here $ v = \sqrt{w^2 + 4 C/w} $ is used as parameter instead of the original strength $C$. Setting
$ \sigma = s^2$ we thus have to minimize
\bea
E_F(v,w) \EA \frac{3}{4} \, \frac{(v-w)^2}{v}  - \frac{2\alpha}{\sqrt{\pi}} \, \frac{v}{w}
\int_0^{\infty} ds \> e^{-s^2} \, \left [ \, 1 + \frac{v^2-w^2}{v w^2} \, \frac{1-e^{-v s^2}}{s^2}
\, \right ]^{-1/2} \non
\EA \frac{3}{4} \, \frac{(v-w)^2}{v}  - \alpha \frac{v}{w} \sum_{n=0}^{\infty} \, b_n(v)
\left ( \frac{v^2-w^2}{v w^2} \right )^n \> ,
\eea
where \cite{fn_10}
\be
b_n(v) \E \frac{2}{\sqrt{\pi}} {-1/2 \choose n} \, \int_0^{\infty} ds \, e^{-s^2} \left (
\frac{1-e^{-v s^2}}{s^2} \right )^n \E \frac{1}{n!} \sum_{k=0}^n {n \choose k} (-1)^k \, \sqrt{ 1 +
k v}^{\, 2k-1} \> .
\ee
For the actual calculation it is more convenient to introduce $ c = v^2/w^2 - 1 = 4 C/w^3 $ so that
\be
E_F(c,v) \E \frac{3}{4} v \left ( 1 - \frac{1}{\sqrt{1+c}} \right )^2 - \alpha \sqrt{1+c} \,
\sum_{n=0}^{\infty} b_n(v) \left ( \frac{c}{v} \right )^n
\ee
and to expand the parameters as
\bea
c \EA c_1 \alpha + c_2 \alpha^2 + \ldots \\
v \EA v_0 + v_1 \alpha + v_2 \alpha^2 + \ldots  \> \> .
\eea
Including terms up to second order in $\alpha$ one finds $v_0 = 3, \, c_1 = 4/27 $, and 
$E_F \to - \alpha - \alpha^2/81 - \ldots $. In higher orders the minimization always leads to 
linear equations for the coefficients $c_n, v_n$ so that they can be solved easily. 
With the help of a symbolic algebra system (such as MAPLE) the higher-order coefficients
$e_n^F$ can then be evaluated in a straightforward manner and are given in 
Eqs. (\ref{e4 Fey}, \ref{e5 Fey}).

It should be noted that in lowest order also the retardation parameter $ w = 3 + {\cal O}(\alpha)$
instead of $ w \to 1 $ as one would have expected naively. This is due to the wrong 
small-$\sigma$ behavior in the ansatz (\ref{f Feyn})
for Feynman's retardation function and would be corrected by an ``improved parametrization''
\cite{WC2}
\be
f_I(\sigma) \E x_I \, \frac{\alpha}{6 \sqrt{\pi}} \> \frac{\exp(-w_I \sigma)}{\sigma^{3/2}} \> .
\label{f imp}
\ee
It is easy to check that both $x_I, w_I \to  1 + {\cal O}(\alpha) $ for small $\alpha$. However,
one can do even better by letting the functional form of the retardation function free. In this
``best quadratic approximation'' \cite{best polaron} one finds
\be
f_{\rm best}(\sigma) \E \frac{\alpha}{6 \sqrt{\pi}} \, 
\frac{\exp(-\sigma)}{[\mu^2_{\rm best}(\sigma)]^{3/2}}
\label{f best}
\ee
for which Eq. (\ref{f imp}) is a convenient approximation since one knows that
generally
\be
\mu^2(\sigma) \> \stackrel{\sigma \to 0}{\longrightarrow} \>  U_0(\sigma) \> \equiv \> \sigma \> .
\label{U_0}
\ee
Indeed, inserting $U_0(\sigma)$ into the virial expression for the polaron ground state energy
\cite{best polaron}
\be
E_{\rm virial} \E - \frac{\alpha}{\sqrt{\pi}} \, \int_0^{\infty} d\sigma \>
\frac{\exp(-\sigma)}{\sqrt{\mu^2(\sigma)}} \, \left ( \frac{3}{2} - \sigma \right )
\label{E virial}
\ee
one obtains $E_{\rm virial} \to -\alpha$ for $\alpha \to 0 $, i.e., $e_1^{\rm best} = -1 $.

In second order we need the first-order change of the profile function and pseudotime
\bea
A_{\rm best}(E) \EA 1 + \alpha \, a_1(E) + \alpha^2 a_2(E) + \ldots
\label{PT profile}\\
\mu^2_{\rm best}(\sigma) \EA \sigma + \alpha U_1(\sigma) + \alpha^2 U_2(\sigma) + \ldots \> \> .
\label{PT amu}
\eea
From the connection (\ref{A from f}) between profile function and retardation function one finds
\be
a_1(E) \E \frac{4}{3 \sqrt{\pi}} \, \int_0^{\infty} d\sigma \> \frac{\exp(-\sigma)}{\sigma^{3/2}} \,
\frac{\sin^2 E\sigma/2}{E^2}
\ee
and therefore from Eq. (\ref{amu from A})
\bea
U_1(\sigma) \EA - \frac{4}{3 \sqrt{\pi}} \, \frac{4}{\pi} \, \int_0^{\infty} d\sigma'
\frac{\exp(-\sigma')}{\sigma'^{3/2}} \, \int_0^{\infty} dE \>
\frac{ \sin^2 E\sigma/2 \sin^2 E\sigma'/2}{E^4} \non
\EA -\frac{1}{9 \sqrt{\pi}} \int_0^{\infty} d\sigma'
\frac{\exp(-\sigma')}{\sigma'^{3/2}} \, \sigma_<^2 \, \left ( 3 \sigma_> - \sigma_< \right ) \> ,
\eea
where $ \sigma_< = {\rm min}(\sigma,\sigma')$. It is possible to express the last integral exactly in
terms of error functions and exponentials. However,
for the calculation of the second-order energy it is better to plug this expression directly into
the virial energy (\ref{E virial}) and expand $ [\mu^2_{\rm best}(\sigma)]^{-1/2} $
up to first order.

Substituting $ \sigma = s^2, \sigma' = s'^2$ we then obtain
\be
E^{\rm best}_0 \E - \alpha - \frac{2 \alpha^2}{9 \pi} \int_0^{\infty} ds \> \frac{3/2-s^2}{s^2} \, 
e^{-s^2} \, \int_0^{\infty} ds' \> \frac{s_<^4 \, (3 s_>^2 - s_<^2)}{s'^2} \,  e^{-s'^2} + 
{\cal O} \left ( \alpha^3 \right )
\ee
where $ s_< = {\rm min}(s,s') $. Introducing polar coordinates $s = r \cos \phi, s' = r \sin \phi$
the integral with $ \pi/4 \le \phi \le \pi/2 $ can be combined with the one in which
$ 0 \le \phi \le \pi/4 $ and one obtains
\be
e_2^{\rm best} \E - \frac{2}{9 \pi} \, \int_0^{\infty} dr \> r^3 \, (3 - r^2) \, e^{-r^2} 
\int_0^{\pi/4} d\phi \> \tan^2 \phi \, \left ( 3 \cos^2 \phi - \sin^2 \phi \right ) 
\E - \left ( \, \frac{1}{12} - \frac{2}{9 \pi} \, \right ) \> .
\label{e2 best}
\ee

\noindent
Higher-order terms may be calculated numerically by using a delay-type equation for the pseudotime
which was found in the variational approximation for worldline QED and dubbed
``variational Abraham-Lorentz equation'' (VALE) \cite{VALE}. It can be easily checked that the
corresponding equation for the three-dimensional polaron case is 
\be
\ddot \mu^2_{\rm best}(\sigma) \> \equiv \> \frac{d^2 \mu^2(\sigma)}{d \sigma^2} 
\E \frac{4}{3} \int_0^{\infty} d\sigma' \> 
\frac{\delta V}{\delta \mu^2(\sigma')}
\, X(\sigma,\sigma') \E \frac{2 \alpha}{3 \sqrt{\pi}} \,  \int_0^{\infty} d\sigma' \>
\frac{\exp(-\sigma')}{ [ \mu^2_{\rm best}(\sigma')]^{3/2}} \, X(\sigma,\sigma') \> ,
\label{polaron VALE}
\ee
where
\be
X(\sigma,\sigma') \Def \mu^2_{\rm best}(\sigma) - \frac{1}{2} \mu^2_{\rm best}(\sigma+\sigma') -  
\frac{1}{2}\mu^2_{\rm best} \left( |\sigma-\sigma' | \right )
\label{def X}
\ee
is the delayed pseudotime (due to the phonon degrees of freedom which have been integrated out).
Equation (\ref{polaron VALE}) may be integrated with the boundary conditions
$ \mu^2(0) = 0 , \dot \mu^2(0) = 1 $ to give
\be
\mu^2_{\rm best}(\sigma) \E \sigma + \frac{2 \alpha}{3 \sqrt{\pi}} \, \int_0^{\sigma} d\sigma' \> 
(\sigma - \sigma')
 \int_0^{\infty} d\sigma'' \>
\frac{\exp(-\sigma'')}{ [ \mu^2_{\rm best}(\sigma'')]^{3/2}} \>  X(\sigma',\sigma'') \> .
\label{amu VALE}
\ee
This gives an iterative scheme to calculate the perturbative terms (\ref{PT amu}) for the pseudotime
and eliminates the corresponding expansion (\ref{PT profile}) for the profile function completely. 
Expanding in powers of $\alpha$ we obtain
\be
U_n(\sigma) \E \frac{2}{3 \sqrt{\pi}} \, \int_0^{\sigma} d\sigma' \> (\sigma - \sigma')
 \int_0^{\infty} d\sigma'' \>'
\frac{\exp(-\sigma'')}{ \sigma''^{3/2}} \> Y_n(\sigma',\sigma') \> , \> \> n \ge 1 \> .
\label{U_n}
\ee
Defining the delayed pseudotime of order $ n $ as
\be
X_n(\sigma',\sigma'') \Def  U_n(\sigma') - \frac{1}{2} \, U_n (\sigma' + \sigma'') - \frac{1}{2} \, 
U_n\left ( |\sigma' - \sigma''| \right ) \> , \hspace{0.3cm} n = 0 , 1, \ldots ,
\ee
the functions $Y_n$ are given by (for simplicity all arguments are suppressed)
\bea
Y_1 \EA X_0 \> , \hspace{0.3cm} Y_2 \E X_1 - \frac{3}{2} \, \frac{X_0 \, U_1}{\sigma''} \> , 
\hspace{0.3cm}
Y_3 \E X_2  - \frac{3}{2}  \, \frac{X_0 \, U_2 + X_1 \, U_1}{\sigma''} +
\frac{15}{8}  \, \frac{X_0 \, U_1^2}{\sigma''^2} \, , \\
Y_4 \EA X_3 -  \frac{3}{2}  \, \frac{X_0  \, U_3 + X_1 \, U_2 + X_2 \, U_1}{\sigma''}
+ \frac{15}{8}  \, \frac{2 X_0 \, U_1 U_2 + X_1 \, U_1^2}{\sigma''^2}
-\frac{35}{16}  \, \frac{X_0 \, U_1^3}{\sigma''^3} \> .
\eea
Once the perturbative terms $U_n(\sigma)$ are known it is straightforward to calculate the energy
coefficients $e_n^{\rm best} , n \ge 1 $ from the virial energy (\ref{E virial})
\be
e_n^{\rm best} \E - \frac{1}{\sqrt{\pi}} \, \int_0^{\infty} d\sigma \> 
\frac{\exp(-\sigma)}{\sqrt{\sigma}}
\, \left ( \frac{3}{2} - \sigma \right ) \> \epsilon_n(\sigma)
\label{en best}
\ee
with (again suppressing the argument $\sigma$)
\bea
\epsilon_1 \EA 1 \> , \hspace{0.3cm}  \epsilon_2 \E - \frac{1}{2}  \, \frac{U_1}{\sigma} \> , 
\hspace{0.3cm}
\epsilon_3 \E - \frac{1}{2}  \, \frac{U_2}{\sigma} + \frac{3}{8} \, \frac{U_1^2}{\sigma^2} \\
\epsilon_4 \EA - \frac{1}{2}  \, \frac{U_3}{\sigma} + \frac{3}{4}  \, \frac{U_1 \, U_2}{\sigma^2}
- \frac{5}{16}  \, \frac{U_1^3}{\sigma^3} \, ,\\
\epsilon_5 \EA - \frac{1}{2}  \, \frac{U_4}{\sigma} + \frac{3}{8}  \,
\frac{2 U_1 \, U_3 + U_2^2}{\sigma^2} - \frac{15}{16}  \, \frac{U_1^2 \, U_2}{\sigma^3}
+ \frac{35}{128}  \, \frac{U_1^4}{\sigma^4} \> .
\label{eps 5}
\eea

We have evaluated Eqs. (\ref{U_n}) - (\ref{eps 5}) by numerical integration. This is a nontrivial task
because of the square-root singularities at $\sigma = 0 $ and the nonanalytic behavior of
$\mu^2 (|\sigma - \sigma'|)$. The first problem was solved by transforming to
$ \sigma = s^2, \sigma' = s'^2 $, etc. , the second one by using the trapezoidal integration rule so 
that $s = s'$ is precisely hit (and not integrated over). In addition, for the first three intervals 
of each integral a Newton-Cotes formula of open type [Eq. 25.4.21 in Ref. \cite{Handbook}] was employed
in order to avoid evaluation of the various integrands at $s = 0$. While this cures the integrable
singularities at the origin, it makes the treatment of the delay more problematic:
in general $ U_n(|\sigma \pm \sigma'| = |s^2 \pm s'^2|) $
is not in the tabulated values of $U_n(\sigma=s^2)$ so that a three-term interpolation had to be  
used. In 
addition, the values of $U_{n>1} $ for small $\sigma$ were determined from the ($\sigma=0$) limit of 
Eqs. (\ref{polaron VALE}), (\ref{def X})
\be
\ddot \mu^2(0) \E - \frac{2 \alpha}{3 \sqrt{\pi}} \, \int_0^{\infty} d\sigma \>
\frac{\exp(-\sigma)}{[\mu^2_{\rm best}(\sigma)]^{1/2}} \> ,
\ee
i.e.,
\be
U_n(\sigma) \> \stackrel{\sigma \to 0}{\longrightarrow} \> \left ( \, - \frac{1}{3 \sqrt{\pi}} \,
\int_0^{\infty} d\sigma \>
\frac{\exp(-\sigma)}{\sqrt{\sigma}} \> \epsilon_n(\sigma) \, \right ) \> \sigma^2  + 
\ldots \> , \hspace{0.3cm} n \ge 1
\ee
with the same functions $\epsilon_n(\sigma)$ as used for calculating the energy coefficients.

Although the trapezoidal (as well as the  Newton-Cotes) integration rule is not very precise [it 
exhibits errors of ${\cal O}(h^3) $ where $ h = s_{\rm max}/N $ is the increment] it offers
an additional advantage: the tabulation of $U_n(\sigma = s^2)$ could be done step by step
avoiding the time-consuming calculation of the integral over $\sigma'$ in Eq. (\ref{amu VALE}) for each
value of $\sigma$. Taking $ \sigma_{\rm max} = 20$ so that the retardation factor $ \exp(-\sigma)$
is sufficiently small at the upper limit of integration, we have achieved stable numerical results
with $ N = 1000 - 1500 $. The numerical value of the second-order coefficent (\ref{e2 best})
was confirmed with high accuracy (seven digits).

\section{Analytical results for the cumulants $\lambda_1$ and $\lambda_2$}
\label{app: m12}
\setcounter{equation}{0}

Here we calculate the cumulants $\lambda_n$ for $ n = 1, 2$.
In the first case $a_{11} = t - t' \equiv \sigma $ and we have for the first moment
\bea
m_1 \EA - \frac{\alpha}{\sqrt{4 \pi}} \, \int_0^{\beta} dt \, \int_0^t dt' \,
\int_0^{\infty} du \>
\frac{\exp(-(t-t'))}{(t-t' + u)^{3/2}} \E -\frac{\alpha}{\sqrt{4 \pi}} \,
\int_0^{\beta} dt \, \int_0^t dt' \, \exp[-(t-t')]
\frac{2}{\sqrt{t-t'}} \non
\EA - \frac{\alpha}{\sqrt{\pi}} \, \int_0^{\beta} d\sigma \,
\int_{\sigma/2}^{\beta - \sigma/2} d\Sigma \>
\frac{\exp(-\sigma)}{\sqrt{\sigma}} \E - \frac{\alpha}{\sqrt{\pi}} \,
\int_0^{\beta} d\sigma \>  \left ( \beta - \sigma \right ) \,
\frac{\exp(-\sigma)}{\sqrt{\sigma}} \> .
\label{m1}
\eea
The remaining $\sigma$ integration is easily done by substituting
$ s = \sigma^2$. This gives 
\be
\lambda_1 \> \equiv \> m_1 \E - \alpha \left [ \, \left (\beta - \frac{1}{2}
\right ) \,
{\rm erf}\left ( \sqrt{\beta} \right ) + \sqrt{\frac{\beta}{\pi}} \, e^{-\beta} \,
\right ] \>
\stackrel{\beta \to \infty}{\longrightarrow}  \>  - \alpha \left [ \, \beta -
\frac{1}{2} + \frac{1}{\sqrt{\beta \pi}} e^{-\beta} + \ldots) \, \right ] \> ,
\ee
where erf$(x)$ is the error function \cite{Handbook}.
Thus we indeed have $e_1 =  -1 $ for the first-order coefficient of the expansion of
the ground-state energy in powers of the coupling constant. It is also seen that the 
subleading term in $\lambda_1$ is a constant which disappears if one calcualates 
the derivative of the cumulant with respect to $\beta$: 
\be
\frac{\partial \lambda_1(\beta)}{\partial \beta} \E - \alpha \, {\rm erf} \left (
\sqrt{\beta} \right ) \>
\stackrel{\beta \to \infty}{\longrightarrow}  \> -  \alpha \left [ \> 1 -
\frac{1}{\sqrt{\beta \pi}} e^{-\beta} + \ldots \> \right ] \> .
\label{asy e1}
\ee

The analytical calculation is more involved for $ n = 2$. We start from Eq. (\ref{diff lambda2 symm})
for the derivative of the second cumulant and substitute $ S = \Sigma_1 - \Sigma_2 $ for the 
integration variable $\Sigma_{1,2} $ with $ \Sigma_{2,1} = \beta - \sigma_{2,1}/2 $ fixed. This gives
\be
\frac{\partial \lambda_2}{\partial \beta} \E  \frac{4 \alpha^2}{ \pi}
\int_0^{\beta} d\sigma_1 \, \int_0^{\sigma_1} d\sigma_2 \, 
\frac{\exp(-\sigma_1 - \sigma_2)}{\sqrt{\sigma_1 \sigma_2}} \int_0^{\beta - s} dS \> f_2 \left (
\frac{a_{12}}{\sqrt{\sigma_1 \sigma_2}} \right ) \> ,
\label{dlambda2}
\ee
where $ s = (\sigma_1+\sigma_2)/2 \ge r = (\sigma_1-\sigma_2)/2 \ge 0 $. The explicit form (\ref{a12}) 
of $a_{12}$ may now be used to write the last integral in Eq. (\ref{dlambda2}) as
\be
\int_0^r dS \>  f_2 \left ( \frac{s-r}{\sqrt{\sigma_1 \sigma_2}} \right ) + 
\int_r^{{\rm min}(s,\beta-2)} dS \>  f_2 \left ( \frac{s-S}{\sqrt{\sigma_1 \sigma_2}} \right ) \> ,
\ee
where the two parts correspond to the constant and linear behavior of $a_{12}$, respectively, 
on the $(S \ge 0)$ side of Fig. \ref{fig: a12}. We thus obtain
\be
\frac{\partial \lambda_2}{\partial \beta} \E  \frac{4 \alpha^2}{ \pi}
\int_0^{\beta} d\sigma_1 \, \int_0^{\sigma_1} d\sigma_2 \, 
e^{-\sigma_1 - \sigma_2} \, \left \{ \, \frac{\sigma_1 - \sigma_2}{\sqrt{\sigma_1 \sigma_2}} \,
f_2 \left ( \sqrt{ \frac{\sigma_2}{\sigma_1}} \right ) + 2
\int_{t_0}^{\sqrt{\sigma_2/\sigma_1}} dt \, f_2(t) \, \right \}
\label{dlambda_2}
\ee
with
\be
t_0(\sigma_1,\sigma_2,\beta) \E \frac{\sigma_1 + \sigma_2 - \beta}{\sqrt{\sigma_1 \sigma_2}} \, 
\Theta(\sigma_1 + \sigma_2 - \beta) \> .
\ee
One sees that for $\beta \to \infty \> \> \> t_0(\sigma_1,\sigma_2,\beta) \to 0 $
since the relative times are bounded by the exponential retardation factors. In other
words
\be
\frac{\partial \lambda_2}{\partial \beta} \To \frac{2 \alpha^2}{ \pi}
\int_0^{\infty} d\sigma_1 \, e^{-\sigma_1} \int_0^{\sigma_1} d\sigma_2 \,
e^{-\sigma_2} \, \left \{ \, \frac{\sigma_1 - \sigma_2}{\sqrt{\sigma_2 \sigma_1}} \,
f_2 \left ( \sqrt{ \frac{\sigma_2}{\sigma_1}} \right ) + 2
\int_0^{\sqrt{\sigma_2/\sigma_1}} dt \, f_2(t) \, \right \}
\ee
and the corrections are of order $ \exp(-\beta) $. Putting $\sigma_2 = t^2 \sigma_1$,
the $\sigma_1$ integration can be performed in the first term and an integration
by parts in the second term gives
\be
\frac{\partial \lambda_2}{\partial \beta} \To \frac{4 \alpha^2}{ \pi}
\, \int_0^1 dt \> \left [ \frac{\arcsin(t)}{t} - 1 \right ] \, \left [
\frac{2}{(1+t^2)^2}
 - \frac{1}{2} \right ] +{\cal O} \left (e^{-\beta} \right) \> .
\ee
Finally a combination of partial integrations [to get rid of the $\arcsin(t)$ ] and
integrals which MAPLE can do, leads to
\be
\frac{\partial \lambda_2}{\partial \beta} \To  \alpha^2 \left [
4 \ln \left ( 1 + \sqrt{2} \right ) - 3 \ln 2 - \sqrt{2} \right ] +
{\cal O} \left (e^{-\beta} \right) \> .
\label{e2(beta)}
\ee
We thus obtain the second-order coefficient of the ground-state energy as given in Eq. (\ref{e2}).
\vspace{0.5cm}

In our approach it is very important to know the precise way how Eq. (\ref{e2(beta)}) approaches the
asymptotic value calculated above. The easiest way to find out is to differentiate
Eq. (\ref{dlambda_2}) again with respect to $ \> \beta$:
\bea
\frac{\partial^2 \lambda_2}{\partial \beta^2} \EA \frac{2 \alpha^2}{ \pi} \, e^{-\beta} \, 
\int_0^{\beta} d\sigma_2 \, e^{-\sigma_2} \, \left \{ \,
\frac{\beta - \sigma_2}{\sqrt{\beta \sigma_2 }} \,
f_2 \left ( \sqrt{ \frac{\sigma_2}{\beta}} \right ) + 
2 \int_{t_0(\beta,\sigma_2,\beta)}^{\sqrt{\sigma_2/\beta}} dt \, f_2(t) \, \right \}
\non
&& -  \frac{4 \alpha^2}{ \pi} \int_0^{\beta} d\sigma_1 \, e^{-\sigma_1}
\int_0^{\sigma_1} d\sigma_2 \, e^{-\sigma_2} \> f_2(t_0) \,
\frac{\partial t_0(\sigma_1,\sigma_2,\beta)}{\partial \beta} \deF I_1 + I_2 \> .
\label{lambda2 ''}
\eea
Consider first the contribution $I_1$: 
since $ \> t_0(\beta,\sigma_2,\beta) \E \sqrt{ \sigma_2/\beta} $ 
is the same as the upper limit of the integral, the latter vanishes so that
\be
I_1 \E \frac{2 \alpha^2}{ \pi} \, e^{-\beta} \, \int_0^{\beta} d\sigma_2 \,
e^{-\sigma_2} \,
\frac{\beta - \sigma_2}{\sqrt{\beta \sigma_2 }} \,
f_2 \left ( \sqrt{ \frac{\sigma_2}{\beta}} \right ) \> .
\ee
The substitution $ \sigma_2 = \beta s^2 $ gives
\be
I_1 \E \frac{4 \alpha^2}{ \pi} \, e^{-\beta} \, \beta \, \int_0^1 ds \,
e^{-\beta s^2} \, \left ( 1 - s^2 \right ) \, \left [ \, \frac{\arcsin s}{s} - 1
\right ]
\ee
and in the limit $\beta \to \infty$ the exponential factor
forces $s \to 0$ in all other terms \cite{fn_11}.
Therefore we may expand these in powers of $s$, integrate
term by term, and obtain
\be
I_1 \> \stackrel{\beta \to \infty}{\longrightarrow} \>   \frac{4 \alpha^2}{ \pi} \,
e^{-\beta} \, \beta \, \int_0^1 ds \,
e^{-\beta s^2} \, \left [ \frac{1}{6} s^2 - \frac{11}{120} s^4 + \ldots \right ] \E
\frac{\alpha^2}{ 6 \sqrt{\pi}} \, \frac{\exp(-\beta)}{\sqrt{\beta}} \,
\left [ \, 1 + {\cal O} \left ( \frac{1}{\beta} \right) \, \right ] \> .
\label{I1}
\ee
For the contribution $ I_2 $ we use 
$ \partial t_0/\partial \beta =  - \Theta(\sigma_1 + \sigma_2 - \beta)/\sqrt{\sigma_1 \sigma_2} $
so that
\be
I_2 \E \frac{4 \alpha^2}{\pi} \,  \int_0^{\beta} d\sigma_1 \, \int_0^{\sigma_1} d\sigma_2 \,
\frac{ \Theta(\sigma_1 + \sigma_2 - \beta) }{\sqrt{\sigma_1 \sigma_2}} \,
e^{-\sigma_1  - \sigma_2} \, f_2 \left (
\frac{ \sigma_1 + \sigma_2 - \beta}{\sqrt{\sigma_1 \sigma_2}} \right ) \> .
\ee
Using the variables of Eq. (\ref{def S,r,s}) we obtain
\be
I_2 \E \frac{4 \alpha^2}{\pi} \, 2 \, \int_{\beta/2}^{\beta} ds \, \int_0^{\beta - s} dr \,
\frac{\exp(-2 s)}{\sqrt{s^2 - r^2}} \, f_2 \left ( \frac{2 s-\beta}{\sqrt{s^2 - r^2}}
\right ) 
\ee
since the Jacobian of the transformation is 2. The
substitutions $ \> s =   \beta (1 + u)/2 , \, r = s  \sin \phi \> $ give
\be
I_2 \E \frac{4 \alpha^2}{\pi} \, \beta \, \beta e^{-\beta} \, \int_0^1 du \, e^{-\beta u} \,
\int_0^{\phi_0(u)} d\phi \>
f_2 \left ( \frac{2 u}{(1+u) \cos \phi} \right ) \> .
\ee
where $ \sin \phi_0(u) = (1-u)/(1+u) $ .
Again, for $\beta \to \infty $ the low-$u$ behavior of the nonexponential part
of the integrand determines the asymptotic behavior.
We have
\be
g(u) \Def \int\limits_0^{\phi_0(u) }d\phi \,
f_2 \left (  \frac{2 u}{1+u} \frac{1}{\cos \phi} \right ) \> \longrightarrow \>
\frac{1}{6} \left ( \frac{2 u}{1+u} \right )^2 \,
\int\limits_0^{\phi_0(u) }d\phi \, \frac{1}{\cos^2 \phi} + \frac{3}{40}
\left ( \frac{2 u}{1+u} \right )^4  \, \int\limits_0^{\phi_0(u) }d\phi \,
\frac{1}{\cos^4 \phi} +  \ldots \> .
\ee
The $\phi$ integrals are elementary (see, e.g., Ref. \cite{Dwight}, pp. 103, 104)
and one obtains
\be
g(u) \> \stackrel{u \to 0}{\longrightarrow} \> \frac{1}{3} u^{3/2} +
{\cal O} \left ( u^{5/2} \right ) \> .
\ee
Therefore
\be
I_2 \> \stackrel{\beta \to \infty}{\longrightarrow} \> \frac{\alpha^2}{\sqrt{\pi}} \, 
\frac{\exp(-\beta)}{\beta^{3/2}}
\ee
is subasymptotic and after integration of Eq. (\ref{I1}) with respect to (large) $\beta$ we 
obtain
\be
e_2(\beta) \> \stackrel{\beta \to \infty}{\longrightarrow} \> e_2 -
\frac{1}{12 \sqrt{\pi}} \, \frac{\exp(-\beta)}{\beta^{1/2}} \> .
\ee
Comparison with Eq. (\ref{asy e1}) shows that this is the same functional
approach to the asymptotic value as for the case $ n = 1 $ ; only the numerical
coefficient is different.

\vspace{1cm}
\section{Tanh-sinh integration}
\label{app: tanh}
\setcounter{equation}{0}

\noindent
Here we briefly outline the ``tanh-sinh integration'' procedure proposed by Takahashi and
Mori \cite{TaMo} and used in most of our deterministic calculations. 
For a one-dimensional integral over the interval $ x \in [-1,+1]$  
it is based on the transformation
\bea
x \EA g(t) \E \tanh \left ( \kappa \, \sinh t \right ) \hspace{1cm} t \in [-\infty,+\infty] \\
g'(t) \EA \frac{1}{\cosh^2 (\kappa \sinh t)} \, \kappa \, \cosh t
\eea
which has the effect that the transformed integrand $ g'(t) \, f(g(t)) $ vanishes at the 
boundaries along with all derivatives [for sufficiently well-behaved $f(x)$]. Therefore
the Euler-Maclaurin summation formula [see, e.g. Ref. \cite{Handbook}, Eq. 25.4.7]
with a stepsize $h$ does not 
get any (power) contributions from the endpoints and we have  
\be
\int_{-1}^{+1} dx \> f(x) \E \int_{-\infty}^{+\infty} dt \> g'(t) \> f \left ( g(t) \right )
\> \approx \> h \,  \sum_{k=-\infty}^{k=+\infty} \> w_k \, f \left ( x_k \right )
\label{trapez}
\ee
with
\bea
x_k \EA g(k h) \equiv  \tanh \left [ \kappa \, \sinh (k h) \right ]
\label{xk}\\
w_k \EA  g'(k h) \equiv \frac{1}{\cosh^2  \left [ \kappa \, \sinh (k h) \right ]} \, 
\kappa \, \cosh (k h) \> .
\label{wk}
\eea
For large $|k|$ and fixed $h$ we find
\bea
x_k \To 1 - 2 \, \exp \left ( - \kappa e^{|k|h} \right ) \> ,
\label{xk large}\\
w_k \To 2 \kappa \,  \exp \left ( - \kappa e^{|k|h} + |k| h \right ) 
\label{wk large}
\eea
showing the ``double-exponential'' character of this transformation.

Although the value $\kappa = \pi/2$ has been reported to be optimal \cite{optimal} we have 
found little difference in efficiency by taking
\be
\kappa \E 1 \> ,
\ee
which is our choice in this work. In practice, the infinite sum in Eq. (\ref{trapez}) is
finite since the weights $w_k$ decrease rapidly with $|k|$ as seen in Eq. (\ref{wk large}).
We use
\be
h |k| \le h \, k_{max} \E 3.4
\label{kmax}
\ee
as a cutoff so that $x_{\pm kmax} = \pm (1 - 2.01 \times 10^{-13})$ and $w_{\pm kmax} 
= 6.02 \times 10^{-12}$.
The number of function calls then is
\be
n_t \E 2 k_{max} + 1 \> .
\ee
Conversely, if $n_t$ is chosen (as we do to estimate the run  time in advance) the 
increment is given by
\be
h \E \frac{6.8}{n_t - 1} \> .
\ee
It is straightforward to extend Eq. (\ref{trapez}) to an arbitrary integral as shown in
Eq. (\ref{tanhsinh a,b}) in the main text.


\begin{thebibliography}{99}
\bibitem{fn_1} One could also consider the perturbative expansion of
the effective mass $m^{\star}(\alpha)$.
\bibitem{fn_2} Except when $\sigma$ is close to $ \beta$ which, however,
is suppressed dynamically by a large spatial difference.
\bibitem{fn_4} Contrary to the mathematical nomenclature we characterize the
principal minors by the index of the row+column which has been cut out.
\bibitem{fn_6} This may be balanced by a better fit to the remaining
data points in which case an unambigous determination of $e_n$ is not possible.
\bibitem{fn_8} This is probably due to an arithmetic error in the fourth cumulant
as Eq. (26) in that paper does not have a proper $\beta \to \infty$ limit while the
lower cumulants in Eqs. (24) and (25) do.
\bibitem{fn_9} It is also the mean value of $ [ \fx(t) - \fx(t') ]^2 $
with respect to the trial action for $\beta \to \infty$
which suggests a more physical interpretation \cite{VALE}.
\end{thebibliography}
\end{document}